\documentclass[twocolumn]{aastex62}

\usepackage{natbib}
\graphicspath{{./}{figures/}}

\submitjournal{ApJ}

\shorttitle{Cas~A Element Asymmetries}
\shortauthors{Holland-Ashford et. al.}

\begin{document}

\title{Asymmetries of Heavy Elements in the Young Supernova Remnant Cassiopeia~A}
\correspondingauthor{Tyler Holland-Ashford}
\email{holland-ashford.1@osu.edu}
\author{Tyler Holland-Ashford}
\affil{Department of Astronomy, The Ohio State University, 140 W. 18th Ave., Columbus, Ohio 43210, USA}
\affil{Center for Cosmology and AstroParticle Physics, The Ohio State University, 191 W. Woodruff Ave., Columbus, OH 43210, USA}

\author{Laura A. Lopez}
\affil{Department of Astronomy, The Ohio State University, 140 W. 18th Ave., Columbus, Ohio 43210, USA}
\affil{Center for Cosmology and AstroParticle Physics, The Ohio State University, 191 W. Woodruff Ave., Columbus, OH 43210, USA}
\affil{Niels Bohr Institute, University of Copenhagen, Blegdamsvej 17, 2100 Copenhagen, Denmark}

\author{Katie Auchettl}
\affil{DARK, Niels Bohr Institute, University of Copenhagen, Lyngbyvej 2, 2100 Copenhagen, Denmark}

\begin{abstract}

Supernova remnants (SNRs) offer the means to study supernovae (SNe) long after the original explosion and can provide a unique insight into the mechanism that governs these energetic events. In this work, we examine the morphologies of X-ray emission from different elements found in the youngest known core-collapse (CC) SNR in the Milky Way, Cassiopeia A.  The heaviest elements exhibit the highest levels of asymmetry, which we relate to the burning process that created the elements and their proximity to the center of explosion. Our findings support recent model predictions that the material closest to the source of explosion will reflect the asymmetries inherent to the SN mechanism. Additionally, we find that the heaviest elements are moving more directly opposed to the neutron star (NS) than the lighter elements. This result is consistent with NS kicks arising from ejecta asymmetries.

\end{abstract}

\keywords{ISM: supernova remnants -- methods: data analysis -- stars: neutron -- supernovae: individual (Cassiopeia~A) -- techniques: imaging spectroscopy -- X-rays: ISM}

\section{Introduction}

In the past decade, 3D simulations of core-collapse supernovae (CCSNe) have improved dramatically. Though the neutrino-driven mechanism was proposed more than five decades ago \citep{colgate66}, it has only begun to produce successful explosions in models in the last few years (see reviews by \citealt{janka16, muller16}). Consequently, simulations are beginning to produce testable predictions of explosion and compact object properties: the distribution of SN energies \citep{muller17a, muller18}, SN light-curves \citep{utrobin17}, nucleosynthetic yields \citep{curtis19}, explosion-generated ejecta asymmetries  \citep{wongwathanarat13, summa18, janka17b} and NS kick velocities \citep{wongwathanarat13,gessner18, janka17,muller18}.

Supernova remnants (SNRs) are a useful sample for comparison to simulation predictions. The heavy elements synthesized in the explosions as well as NSs are observable in young SNRs of ages $\lesssim10^{4}$~years (see reviews by \citealt{weiss06b} and \citealt{vink12}). As the ejecta expands into the interstellar medium (ISM), the reverse shock heats the ejecta to $\sim10^{7}$~K temperatures, producing X-rays that can be detected with modern X-ray facilities, such as the {\it Chandra} X-ray Observatory. X-ray studies of supernova remnants can reveal the presence and properties of shocked ejecta (e.g., \citealt{hwang08, lopez09b, luna16, bhalerao19}). Optical and infrared studies are also a useful means to probe the three-dimensional structure of ejecta (e.g., \citealt{reed95, fesen01,gerardy01,delaney10,milisavljevic13,bonchul18}).

Cassiopeia~A (Cas~A hereafter) is a prime target for comparison to current CC SNe simulations as it is the youngest known CC SNR in the Milky Way ($\approx$350 years old; \citealt{thorstensen01}). Cas~A's X-ray emission is dominated by the ejecta metals, and its proximity (with a distance of 3.4~kpc; \citealt{fesen06}) enables investigation of the distribution of metals on small (sub-parsec) scales. Optical observations have shown that Cas~A is an O-rich SNR (e.g., \citealt{chevalier78}), and light echoes from the explosion reveal that Cas~A was produced by an asymmetric Type IIb SN with variations in ejecta velocities of $\approx$4000 km s$^{-1}$ \citep{rest11}. Cas~A was the target of {\it Chandra}'s first light image \citep{hughes00}, and since then, {\it Chandra} has observed the SNR for $\sim$3 Ms. 

Prominent features of Cas~A are its distinct, fast-moving ejecta knots \citep{fesen06, delaney10, milisavljevic15} which span from the center of the SNR to beyond the forward shock \citep{hughes00,hwang03}, thin synchrotron filaments around its periphery \citep{gotthelf01, vink03}, a NS \citep{tananbaum99}, and bright X-ray emission from intermediate-mass elements (Si, S, Ca, Ar) and heavy elements (Ti and Fe; \citealt{vink96}). Based on a detailed analysis of the deep {\it Chandra} data, \cite{hwang12} constructed maps of the elemental abundances, emission measures, and plasma ionization states across Cas~A. Recent simulations \citep{orlando16, janka17b, wongwathanarat17} have aimed to reproduce the observed characteristics of Cas~A, such as the jet (e.g., \citealt{fesen16}), the heavy element abundances and spatial distribution (e.g., \citealt{hwang12, grefenstette17}), and the NS kick velocity (e.g., \citealt{thorstensen01}). 

In this paper, we use the available {\it Chandra}  and {\it NuSTAR} observations of Cas~A to measure the asymmetries of elements heavier than carbon and compare the results to the simulation predictions that heavier elements are ejected more asymmetrically and more directly opposed to NS motion than lighter elements \citep{wongwathanarat13, janka17,gessner18,muller18}). Though recent work by \cite{me17} and \cite{katsuda18} showed that NSs are preferentially kicked opposite to the bulk of ejecta in several young Galactic SNRs, to date, no studies have compared the relative symmetries of different elements within individual SNRs and how those asymmetries relate to the NS motion.

This paper is structured as follows. In Section~\ref{sec:data}, we describe the observations analyzed in this study. In Section~\ref{sec:methods}, we outline the methods used to measure the X-ray morphology and the details of the spectral analysis performed. In Section~\ref{sec:results}, we present our results. Section~\ref{sec:conc} summarizes our conclusions and outlines possible future work.  

\section{Data and Spectral Analysis} \label{sec:data}

For our analysis, we use 15 archival {\it Chandra} X-ray observations of Cas~A (see Table~\ref{table:obslog}) from the Advanced CCD Imaging Spectrometer (ACIS), totaling $\sim$1.3 Ms of exposure time, with $\sim$3$\times$10$^8$ counts in the full (0.5--8.0~keV) band. Cas~A, with a diameter of $\approx$6\arcmin, fits on the ACIS-S3 chip. Although more data were available (Cas~A has been observed for a total of $\sim$3 Ms), we found that spectra had sufficient signal-to-noise in the 15 longest ACIS observations to fit the spectra. Inclusion of the shorter ACIS observations dramatically increased the computation time to fit spectra, and thus we opted to limit our analysis to the observations in Table~\ref{table:obslog}.

\begin{deluxetable}{lccc}
\tablecolumns{4}
\tablewidth{0pt} 
\tablecaption{{\it Chandra} {\sc ACIS} Observations\label{table:obslog}} 
\tablehead{ \colhead{ObsID} & \colhead{Date} &\colhead{Roll} &\colhead{Exposure}  \\
\colhead{} &\colhead{} &\colhead{Angle ($\deg$)} & \colhead{(ks)}  }
\startdata
1952 & 2002-02-06 & 323.4 & 50 \\
4634 & 2004-04-28 & 59.22 & 148 \\
4635 & 2004-05-01 & 59.22 & 138 \\
4636 & 2004-04-20 & 49.77 & 150 \\
4637 & 2004-04-22 & 49.77 & 170 \\
4638 & 2004-04-14 & 40.33 & 170 \\
4639 & 2004-04-25 & 49.77 & 80 \\
5196 & 2004-02-08 & 325.5 & 50 \\
5319 & 2004-04-18 & 49.77 & 40 \\
5320 & 2004-05-05 & 65.14 & 54 \\
10936 & 2010-10-31 & 236.5 & 31 \\
14229 & 2012-05-15 & 75.44 & 50 \\
14480 & 2013-05-20 & 75.14 & 50 \\
14481 & 2014-05-12 & 75.14 & 50 \\
14482 & 2015-04-30 & 67.13 & 50
\enddata
\end{deluxetable}

To study the distribution of Ti, we used the 4.6 Ms background-subtracted 65--70 keV image of Cas~A taken with {\it NuSTAR}  \citep{grefenstette14, grefenstette17}, which is dominated by the radioactive decay line of $^{44}$Ti at 67.87~keV. These data were taken from August 2012 to December 2013, using 14 combined FPMA$+$FPMB images (see \citealt{grefenstette17} for details of the observations and data reduction). Using the spatially-resolved spectral fits to the $^{44}$Ti line and non-thermal continuum reported in Table~2 of \cite{grefenstette17}, we estimate that the $^{44}$Ti line produces 80--100\% of the flux in the 65.0--70.0~keV band. Thus, the 65--70~keV morphology mostly reflects the distribution of the radioactive Ti in the SNR.

Our goal is to measure the morphologies of metals synthesized in the explosion. Although narrow-band images can be produced by filtering to the energy ranges that correspond to prominent X-ray emission lines (see Table~\ref{table:lines}), the resulting images would include the continuum emission (bremsstrahlung and synchrotron) in those bandpasses and not solely reflect line emission from a specific element. Thus, we aim to remove the continuum emission by performing a spatially-resolved spectral analysis. Specifically, we divide the SNR into 2517 regions, model each region's spectra, and subtract the continuum emission in each region from the narrow-band images. The resulting continuum-subtracted images should contain flux only from the emission lines of the elements in question: O, Mg, Si, S, Ar, Ca, Fe. Furthermore, we then convert these element-specific flux images into emission measure maps, dividing flux by the emissivity calculated using spectrally-fit parameters such as temperature and ionization timescale (see Section~\ref{subsec:masscalc}). As we do not expect the continuum to be a dominant contributor at the energies of the Ti line, we do not make continuum-subtracted images or mass maps for our {\it NuSTAR} observation. As such, we also do not convert the Ti image to emission measure maps.

\begin{figure*}
\begin{center}
\includegraphics[width=0.93\columnwidth]{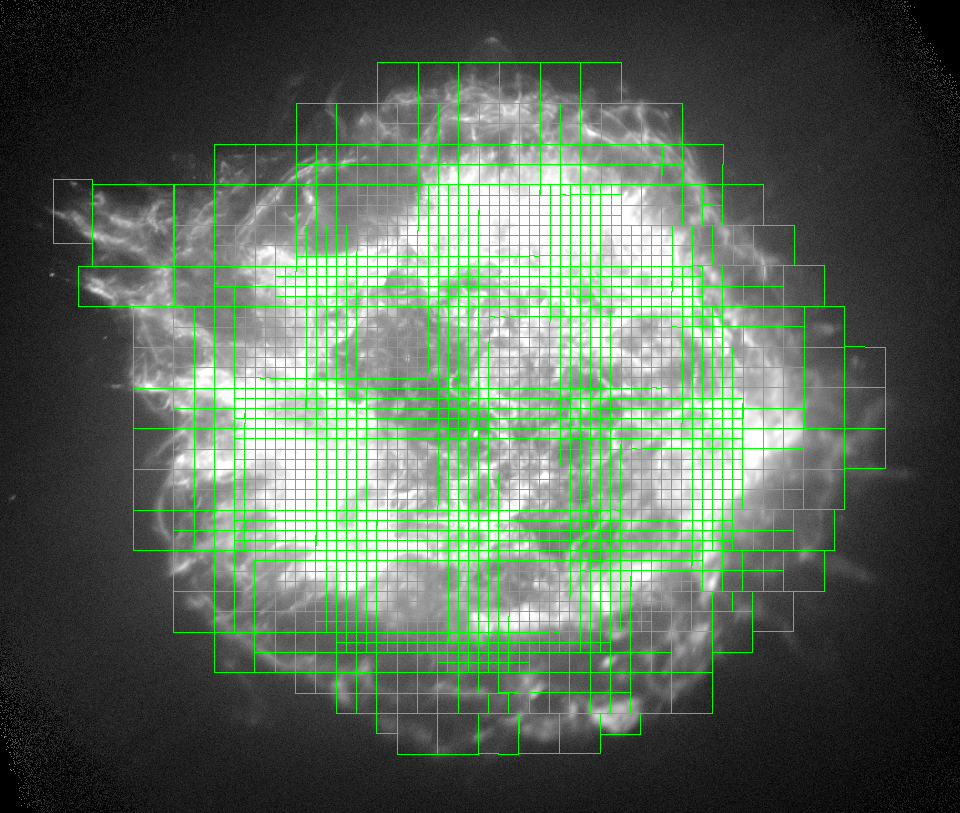}
\includegraphics[width=1.07\columnwidth]{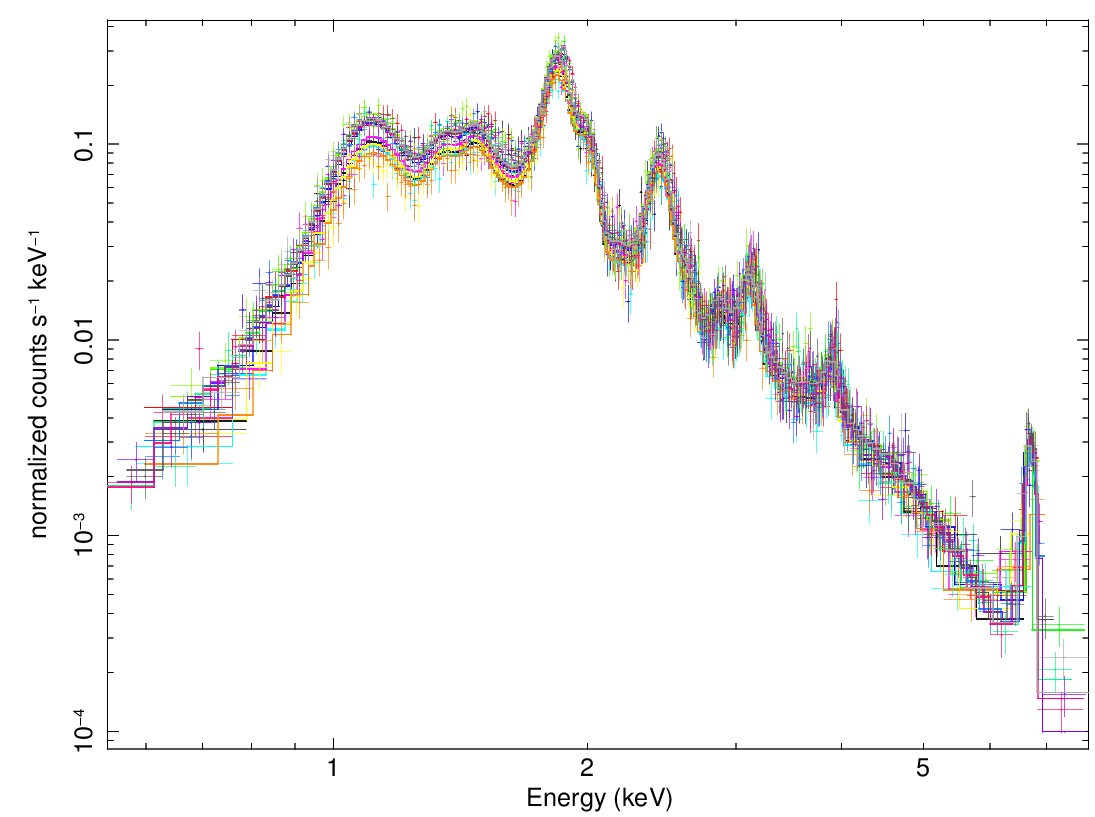}
\end{center}
\caption{{\it Left:} the full-band image of Cas~A with an overlay of the regions analyzed. {\it Right:} a sample fit spectra with a strong Fe-K feature, taken from the bright, eastern region of Cas~A. In this specific example, a two-component model was used a the single-component model did not capture the Fe line.}
\label{fig:grid}
\end{figure*}

\subsection{Narrow-Band Images}\label{subsec:narrowband}

To produce the narrow-band images corresponding to emission lines from each element (see Table~\ref{table:lines} for energy ranges), we used the {\it Chandra} Interactive Analysis of Observations ({\sc ciao}) Version 4.7 and {\sc ftools} \citep{blackburn95}. Specifically, we produced exposure-corrected, merged {\it Chandra} images from the 15 observations using the {\sc ciao} command {\it merge\_obs}. To remove the NS from the resulting image, we used the {\sc ciao} command {\it wavdetect} to define a region around the source, and then used the command {\it dmfilth} to replace this region with values interpolated from surrounding regions. We note that narrow-band analysis is a common technique used to measure ejecta distributions (e.g., \citealt{hwang00, park02, alan19}). Narrow-band images are often improved by subtracting a continuum component -- created using interpolations from energy bands dominated by continuum -- and producing line-to-continuum ratio images that reveal regions of enhanced line emission. In this paper, given the deep observations of Cas~A now available, we evaluate the line emission through detailed spatially-resolved spectral analysis.

\subsection{Spectral Grid}\label{subsec:grid}

For our spatially-resolved spectral analysis, we first split the remnant into small boxes, of sizes 5\arcsec$\times$5\arcsec, 10\arcsec$\times$10\arcsec, or 20\arcsec$\times$20\arcsec (see Figure~\ref{fig:grid}). The smallest boxes were used for regions inside the SNR's contact discontinuity, within 130\arcsec\ of the SNR's center \citep{gotthelf01, bleeker01}, where signal was sufficient to model the spectra robustly. The 10\arcsec$\times$10\arcsec\ boxes spanned between the contact discontinuity and the forward shock (at $\sim$160\arcsec), and the 20\arcsec$\times$20\arcsec\ boxes were employed at larger radii. For the regions around the northeast jet, we aligned the boxes manually to match its profile. This method ensured that each box, except for the most outer regions, contained $>$10,000 counts. Typical values of signal-to-noise were $\sim$200, reaching up to $\sim$700 for the 5\arcsec$\times$5\arcsec boxes with the strongest signal and down to $\sim$60 in the most faint 20\arcsec$\times$20\arcsec boxes. Our regions had reduced chi-squared values of 1.5-2. In addition, our minimum box size of 5\arcsec$\times$5\arcsec\ is much larger than the {\it Chandra} ACIS PSF of 0.5\arcsec, ensuring that each box contains distinct signal. We note that for an off-axis angle of 5\arcmin, (the diameter of Cas~A) the PSF of {\it Chandra} is still $\lesssim$ 5\arcsec, our minimum region size.\footnote{http://cxc.harvard.edu/proposer/POG/}. Thus, the differences in roll angle between the various images should not affect the spectral analysis.

Following this procedure, we extracted spectra from 2517 regions from all 15 {\it Chandra} observations (see Table~\ref{table:obslog}), binning to 20 counts per bin and subtracting the background obtained from a region outside of the SNR. We note that in certain locations, the data has sufficient counts to perform spectral analysis in regions of 2.5\arcsec$\times$2.5\arcsec\ or smaller. However, upon examination, we found that no significant changes in best-fit parameters occurred between the spectra from the 2.5\arcsec$\times$2.5\arcsec\ versus the 5\arcsec$\times$5\arcsec\ regions. Thus, to keep the analysis manageable, we chose to adopt 5\arcsec$\times$5\arcsec\ regions as the smallest box size. We investigated these boxes with the highest signal ($\gtrsim$200,000 counts, S/N $\gtrsim$400) to check that the fits were formally acceptable, and found that the fits were good with typical reduced chi-squared values of $\sim$1.6.

\subsection{Spectral Fitting}\label{subsec:spectral}

We simultaneously fit the 15 spectra from each region using XSPEC Version 12.9.0 \citep{arnaud96} with AtomDB\footnote{atomdb.org} 3.0.7 \citep{smith01,foster12}. In a first pass through the data, we modeled every spectrum as an absorbed ({\it phabs}) thermal, non-equilibrium ionization (NEI) plasma  ({\it vpshock}) plus a power-law component (with the photon-index fixed to 2.5 to model the synchrotron emission). We let the ionization timescale of the {\it vpshock} component vary up to $10^{12}$ cm$^{-3}$~s, which would reflect a plasma in collisional ionization equilibrium (CIE, \citealt{smith10}). Past work has demonstrated that the plasma in Cas~A is in a NEI state \citep{markert82, hwang12,rutherford13}, consistent with our best-fit ionization timescales. As oxygen is the primary source of the thermal continuum \citep{chevalier78, vink96, laming03, hwang03, hwang12}, we adopted an oxygen abundance of 1 relative to solar and allowed the heavier elements (Mg, Si, S, Ar, Ca, Fe, Ni) to vary freely, assuming solar abundances from \cite{anders89}.  

\begin{figure*}
 \includegraphics[width=0.24\textwidth]{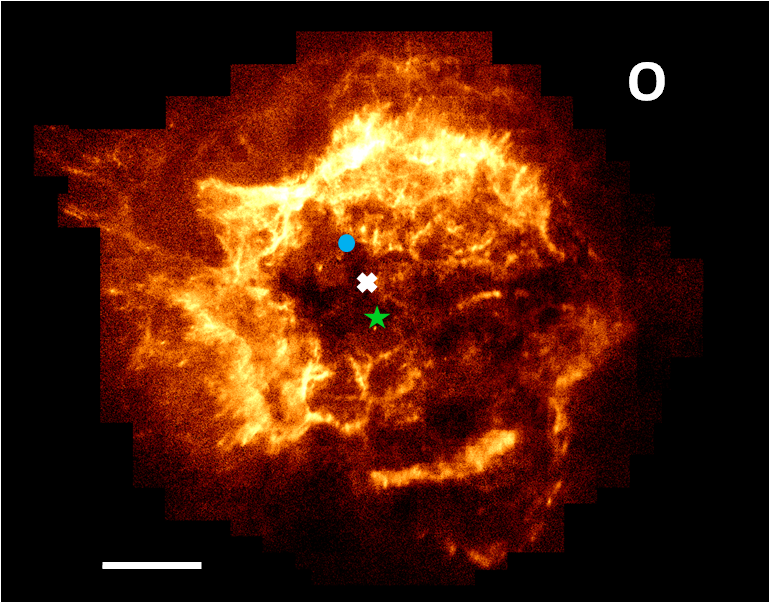}
 \includegraphics[width=0.24\textwidth]{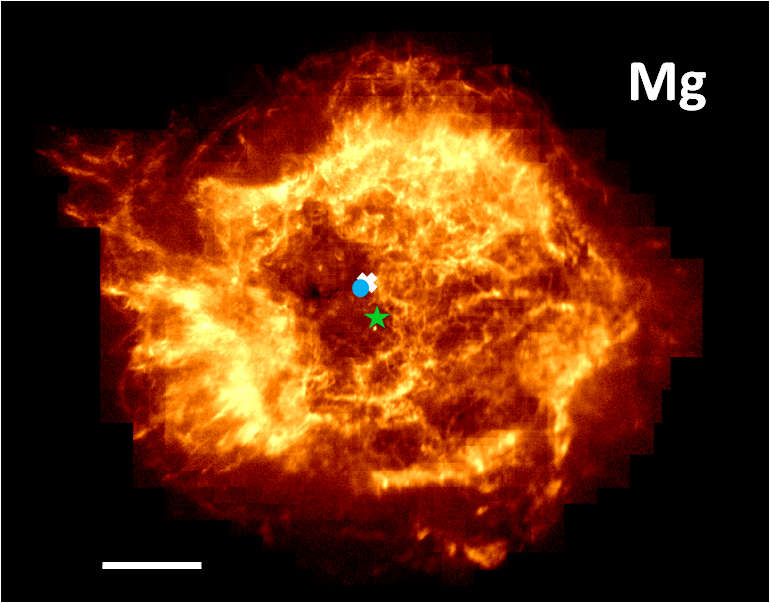}
 \includegraphics[width=0.24\textwidth]{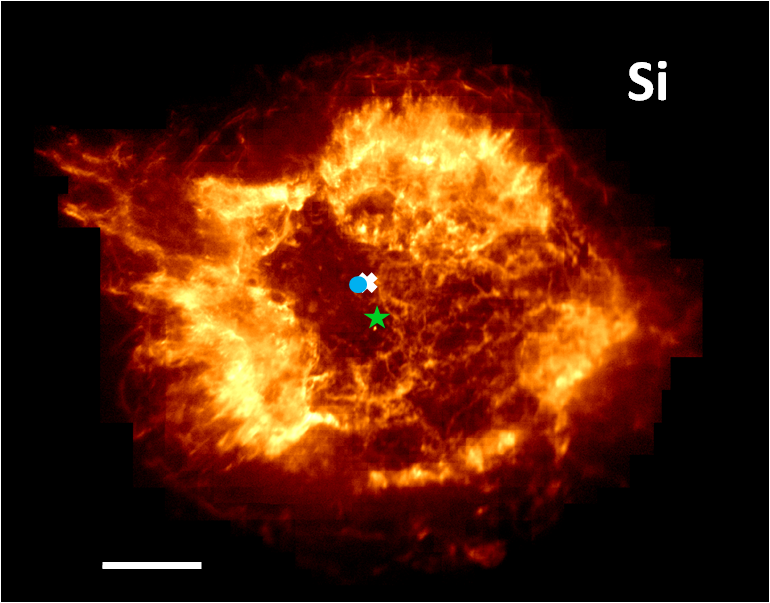}
  \includegraphics[width=0.24\textwidth]{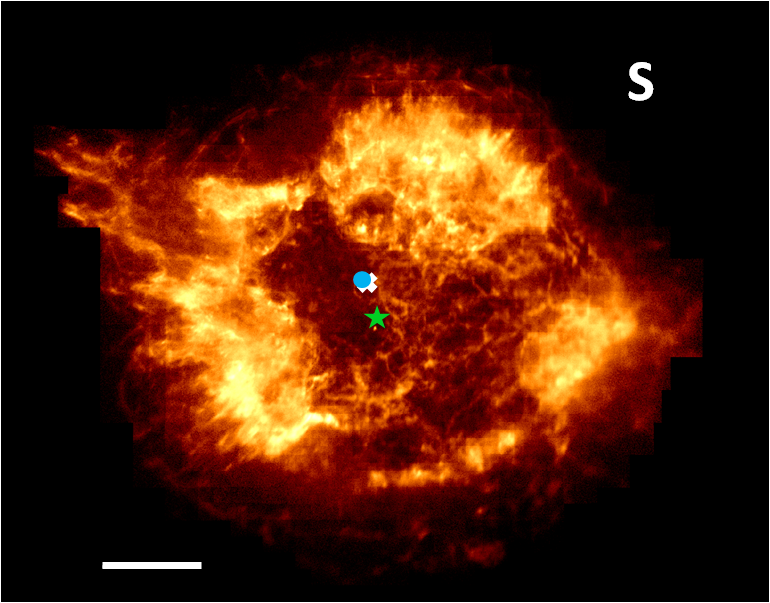}
 \\
 \includegraphics[width=0.24\textwidth]{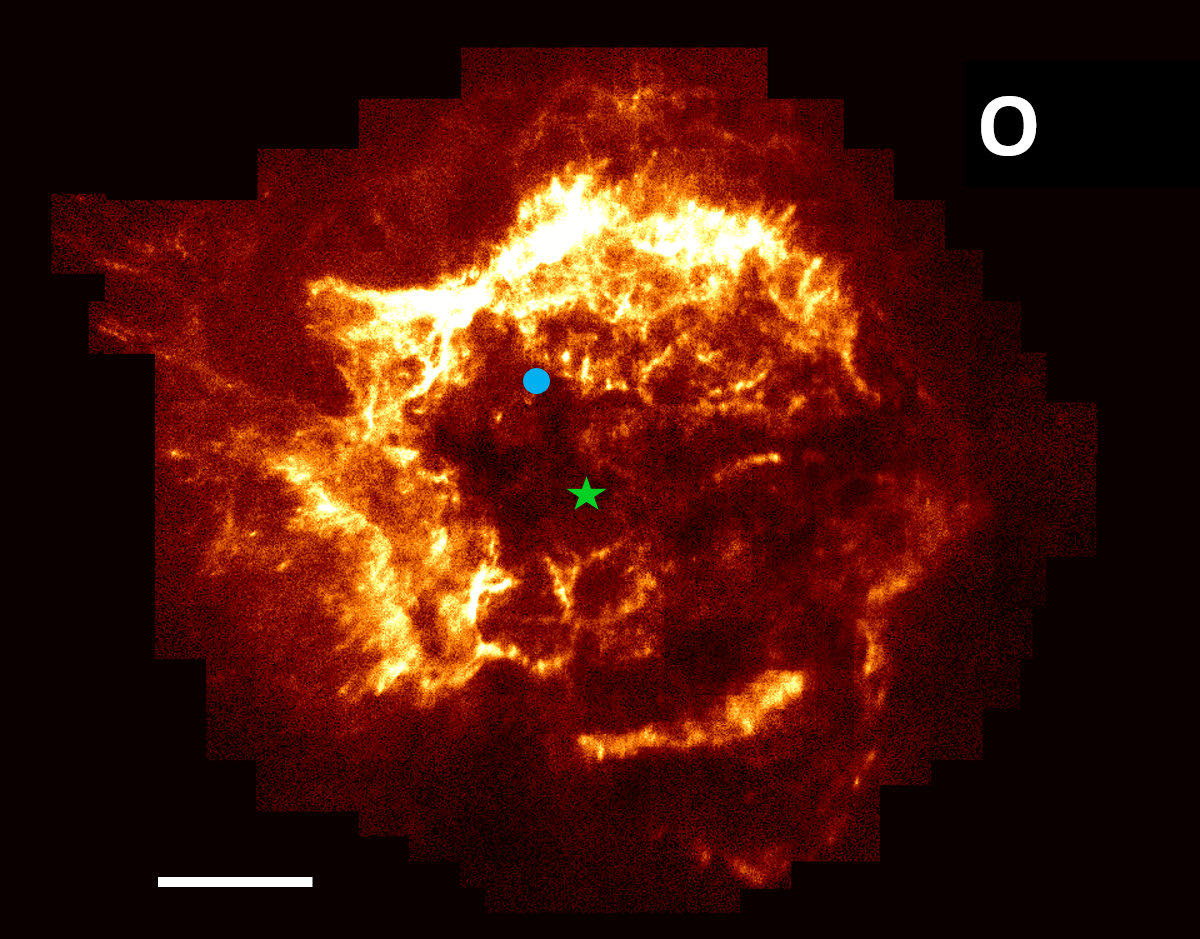}
 \includegraphics[width=0.24\textwidth]{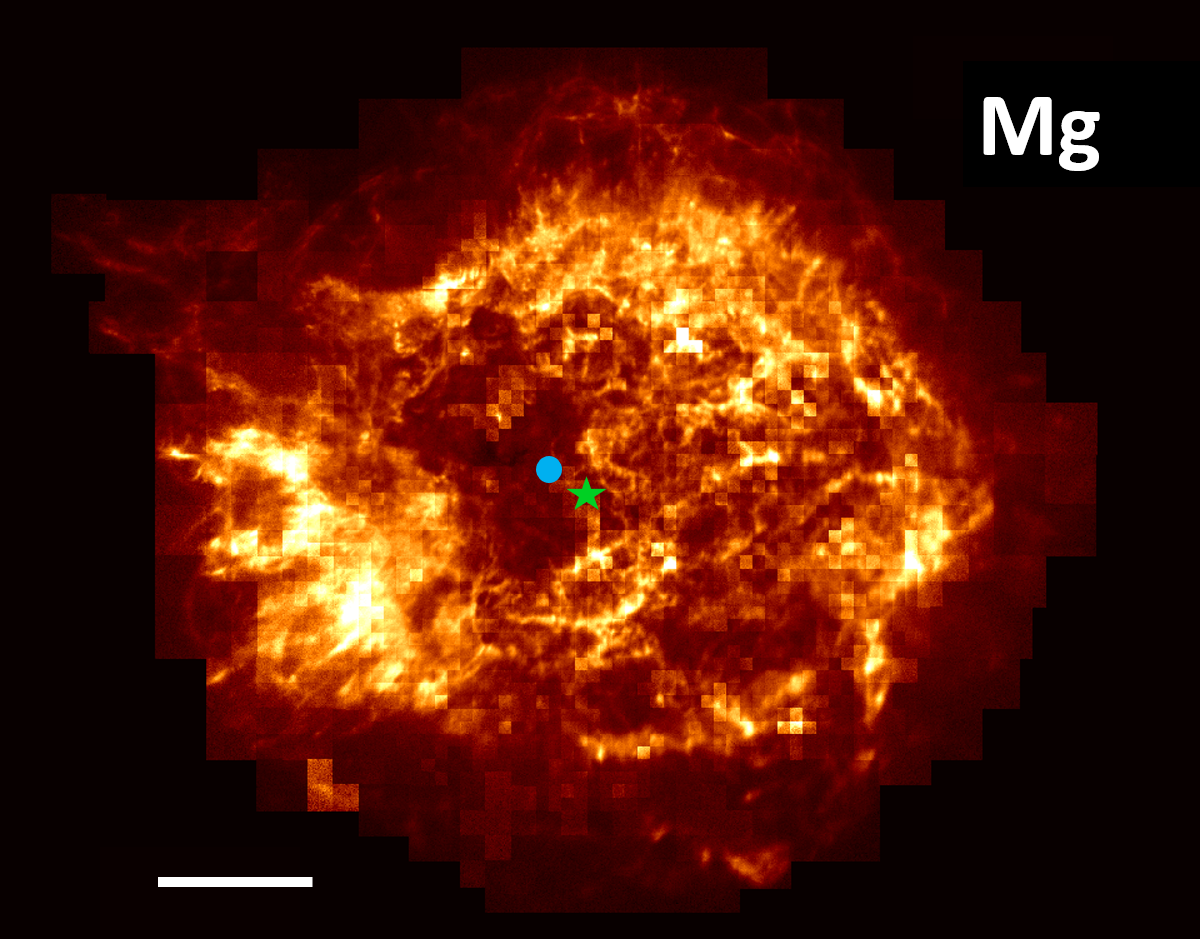}
 \includegraphics[width=0.24\textwidth]{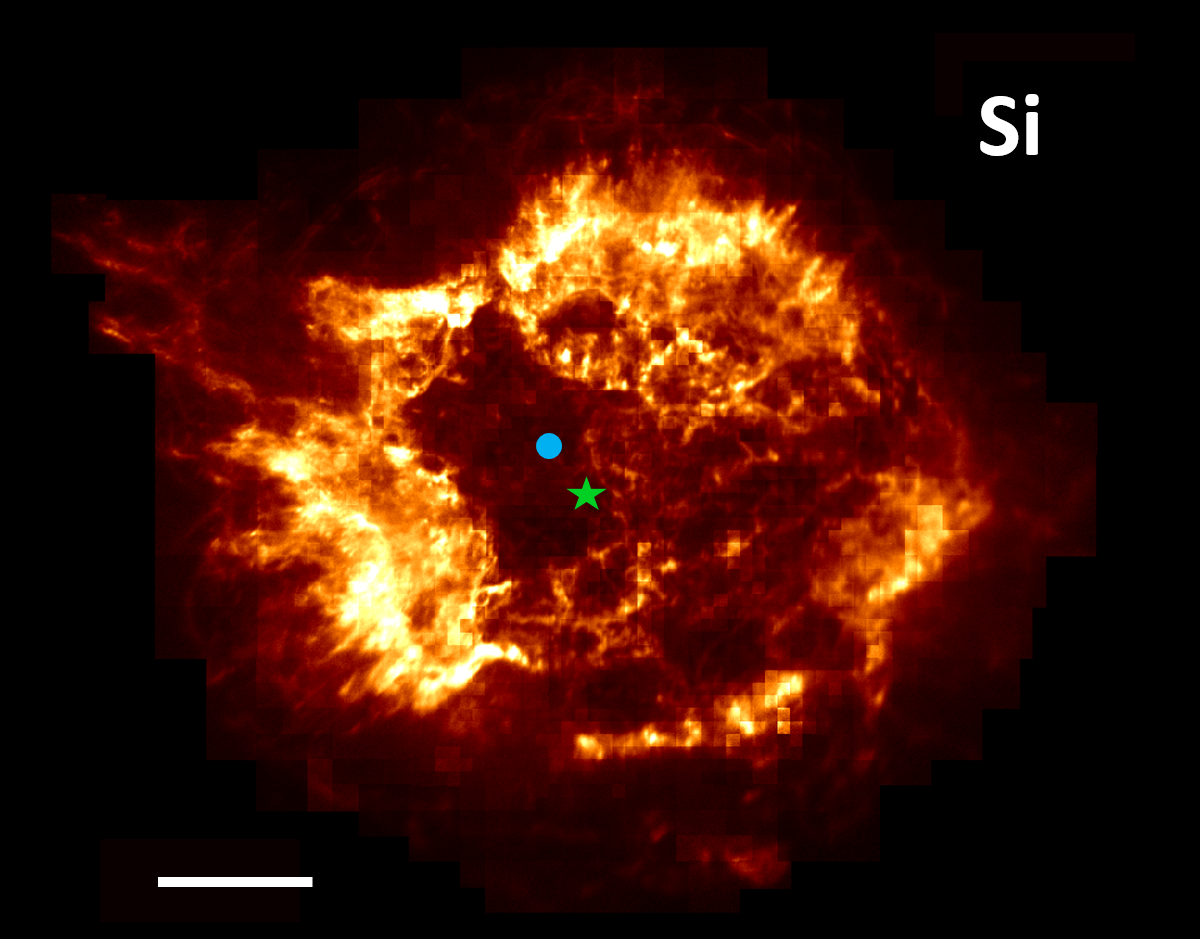}
  \includegraphics[width=0.24\textwidth]{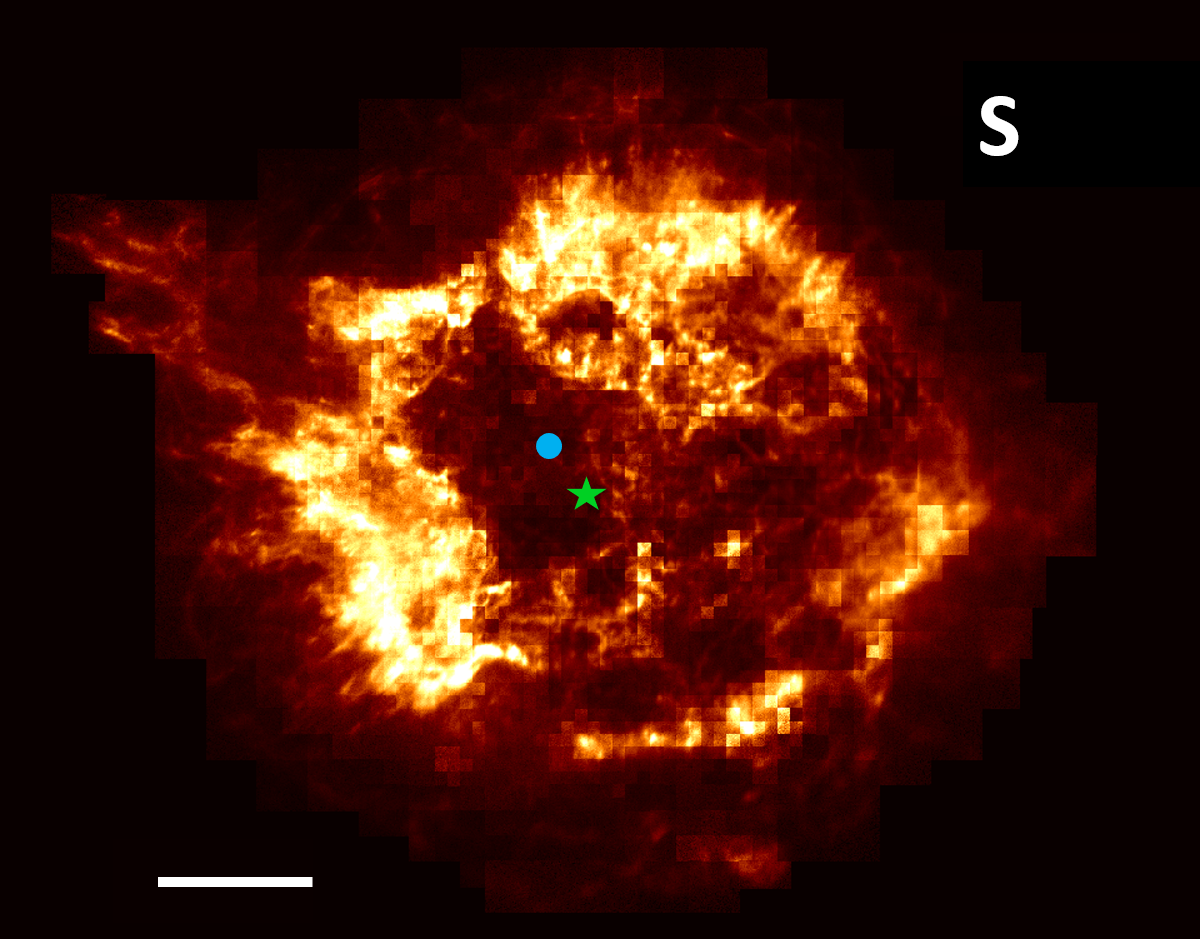}
 \\
 \\
 \includegraphics[width=0.24\textwidth]{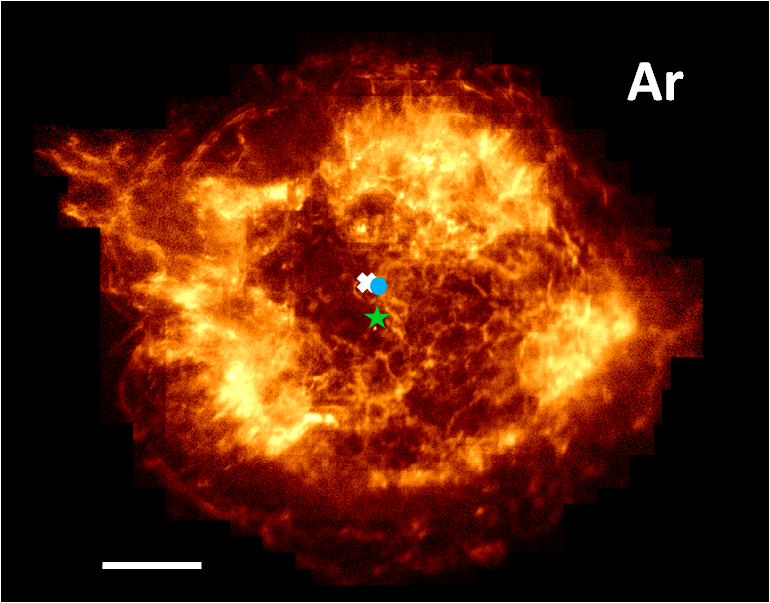}
 \includegraphics[width=0.24\textwidth]{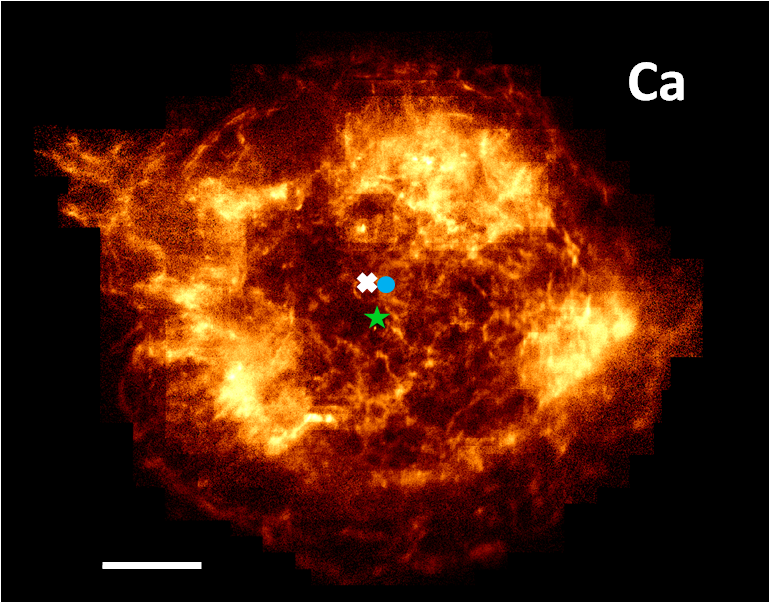}
 \includegraphics[width=0.24\textwidth]{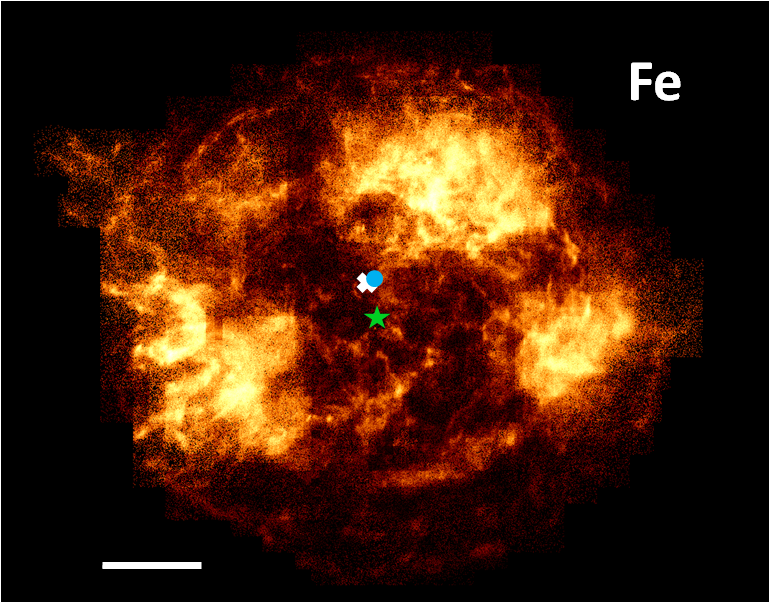}
  \\
 \includegraphics[width=0.24\textwidth]{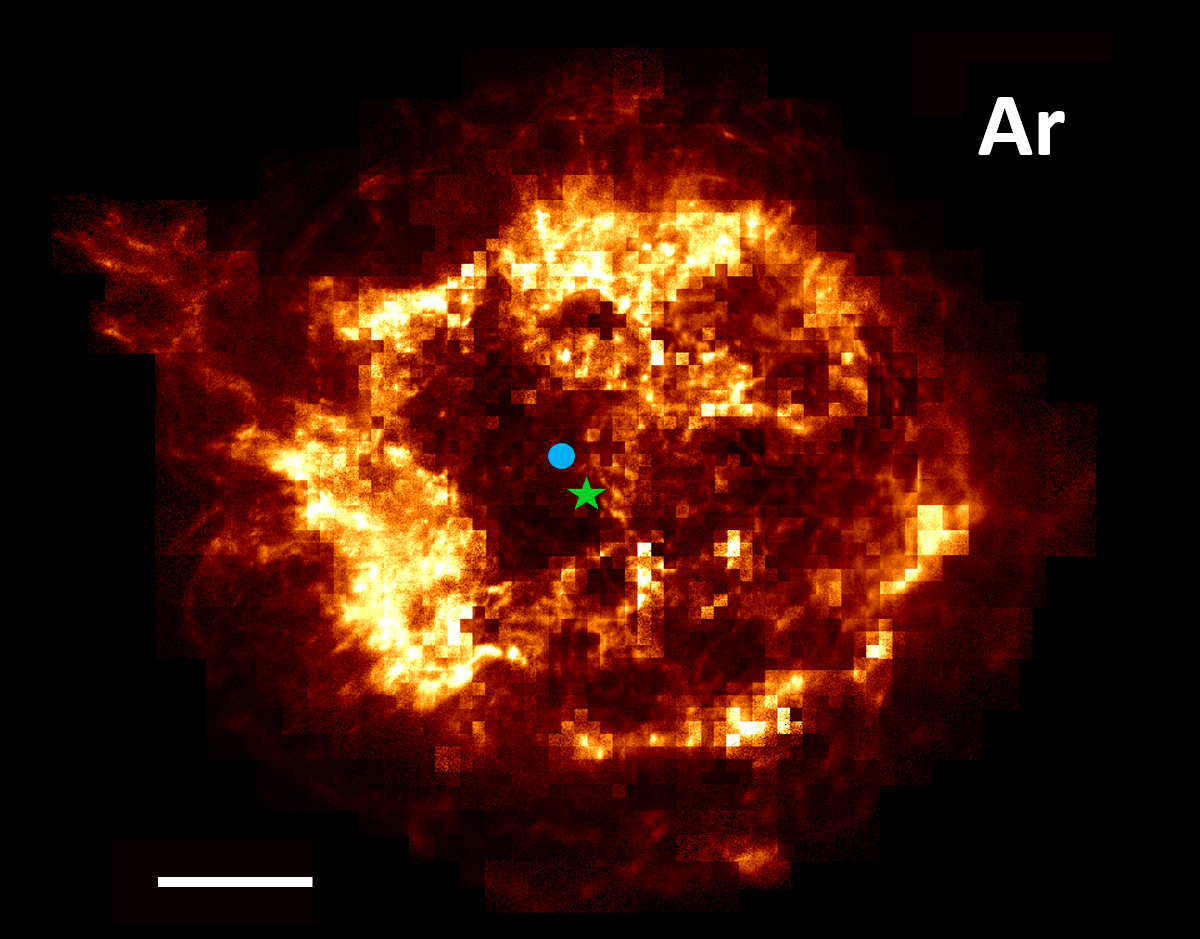}
 \includegraphics[width=0.24\textwidth]{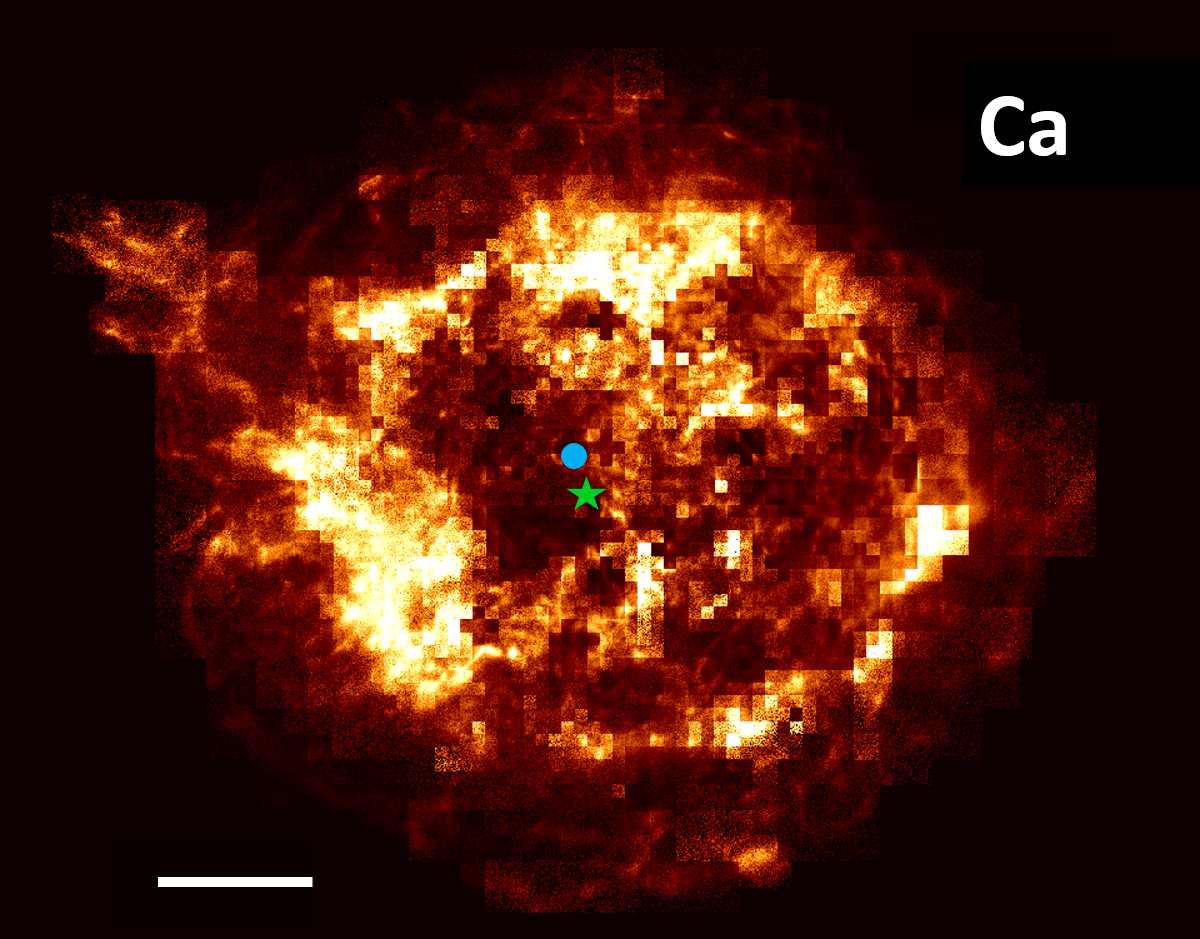}
 \includegraphics[width=0.24\textwidth]{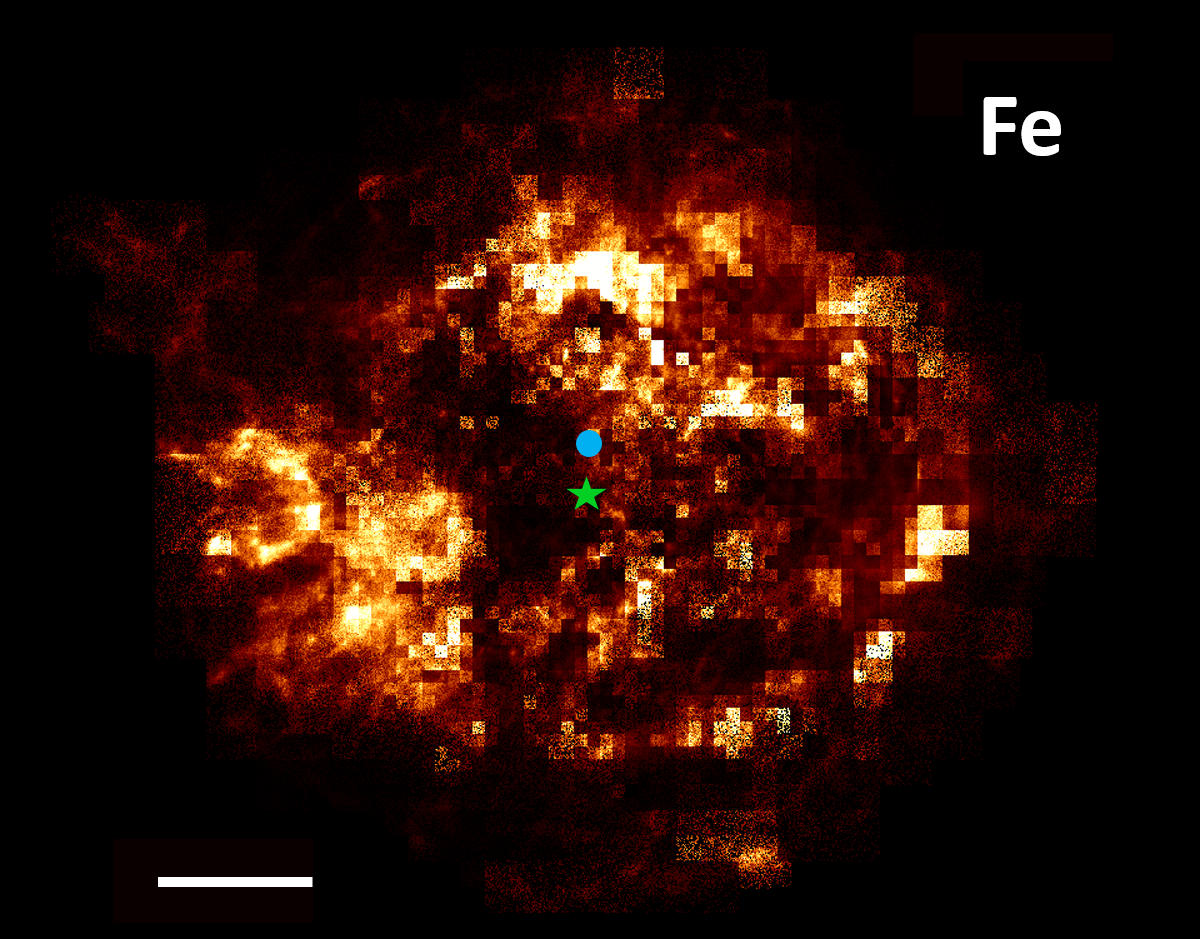}

 \caption{Continuum-subtracted images (1st and 3rd rows) and emission measure maps (2nd and 4th rows) of elements in Cas~A. For the continuum-subtracted images, the green star is the explosion site \citep{thorstensen01}, the white `X' is the full-band (0.5--8.0~keV) center-of-emission, and the blue circle is the center-of-emission for the element displayed. For the emission measure maps, the green star is the explosion site and the blue circle is the center of the emission measure, a proxy for the center-of-mass. The white scale bar at the bottom-left of each image is 2\arcmin\ in length. See \cite{grefenstette17} for the narrow-band Ti image we used.}
  \label{fig:images} 
 \end{figure*}

We fit the spectra, initially allowing only the column density $N_{\rm H}$, temperature $kT$, ionization timescale, and normalization to vary. Subsequently, we thawed the elemental abundances and refit. Finally, to account for the Doppler shifts of the expanding ejecta, we allowed the redshift to vary and refit once more. This procedure was chosen after performing tests on spectra of different signal-to-noise levels and spectral shapes, and we found that it was the most successful at producing accurate spectral fits that did not get stuck in $\chi^{2}$ local minima. After these steps, if a region's fit yielded a high reduced chi-squared ($\gtrsim$4) or if the model did not account sufficiently for the Fe K line, we added a second {\it vpshock} component with a greater plasma temperature. For this second component, we found that allowing the temperature and Fe abundance to vary (with all the other abundances set to solar) was sufficient to reproduce the spectra and reflects the Fe-rich ejecta identified by \cite{hwang12} and \cite{katsuda18}.

In Figure~\ref{fig:grid} we plot an example of the best-fit, two {\it vpshock} $+$ power-law model used to reproduce the emission from a region on the eastern side of the SNR. We note that \cite{hwang12} also found that a model with two thermal components ({\it vpshock}+NEI) was needed to accurately fit the Fe K lines in many regions. They adopted a single ionization timescale (of 8 $\times$ 10$^{11}$ cm$^{-3}$ s) and temperature (1.95 keV) to match the values of the most enriched Fe ejecta spectra from \cite{hwang03}, while we allowed these parameters to vary. We get similar results to those of \citet{hwang03}, and in Section~\ref{subsec:hwang}, we present a more detailed comparison between our results.

We recognize that single- or two-component models are likely simplified approximations of the true state of the SNR plasma. In reality, each region may have signal from plasma of varied thermodynamic conditions (e.g., multiple temperatures, ionization timescales, abundances). However, we found that the single- or two-component models yielded reasonable $\chi^2$ fits, so for simplicity, we opted not to include additional components.

\begin{deluxetable}{ccrccc}
\tablecolumns{5}
\tablewidth{0pt} 
\tablecaption{Metals and X-ray Lines\label{table:lines}} 
\tablehead{ \colhead{Element} & \colhead{Line} & \colhead{Bandpass} & \colhead{Burning}  & \colhead{NS Angle\tablenotemark{c}} & \colhead{NS Angle\tablenotemark{d}} \\
\colhead{} &\colhead{} & \colhead{(keV)\tablenotemark{a}} & \colhead{Process\tablenotemark{b}} & \colhead{(degrees)} & \colhead{(degrees)}}
\startdata
O& VII & 0.55--0.75 & He & 146 & 146\\
Mg& XI & 1.25--1.45 & C & 141 & 105\\
Si& XIII & 1.70--2.1& O, xO & 140  & 128\\
S& XV & 2.25--2.95 & O, xO & 148  & 127\\
Ar& XVII & 3.00--3.25 & O, xO & 172  & 133\\
Ca& XIX & 3.80--4.00 & O, xO & 174 & 141\\
Fe& XXV & 6.35--6.90 & xSi & 165 & 168\\
Ti& decay\tablenotemark{e} & 65--70 & xSi & 160 & N/A \\
\enddata
\tablenotetext{a}{Bandpasses were selected by visual inspection of lines in multiple of our spectra.}
\tablenotetext{b}{The primary burning process that creates each element. An ``x'' in front of an element signifies explosive burning of that element.}
\tablenotetext{c}{Angle between the NS direction of motion and each element's center-of-emission.}
\tablenotetext{d}{Angle between the NS direction of motion and each element's center-of-mass.}
\tablenotetext{e}{$^{44}$Ti radioactively decays into $^{44}$Sc, which then further decays into $^{44}$Ca, to produce the 67.87 keV line \citep{chen11}.}
\end{deluxetable}

\subsection{Continuum-Subtracted Images}\label{subsec:contsub}

To produce continuum-subtracted images, we modeled just the thermal$+$non-thermal continuum in each region as well. To model the thermal continuum emission, we used the AtomDB NoLine\footnote{http://www.atomdb.org/noline.php} model ({\it apecnoline}), using the same $N_{\rm H}$, $kT$, normalization, and redshift values calculated for the {\it vpshock}+power-law model. To model the non-thermal component, we simply used aother power-law with the same normalization and photon-index as calculated for the {\it vpshock}+power-law model. Added together, these models reflected the total thermal$+$non-thermal continuum in each region.
We then derived the fraction of emission arising from the corresponding element's emission line, $F_{\rm elem}$, given by

\begin{equation} \label{eq:pelem}
F_{\text{elem}} = 1-\frac{\text{continuum flux}}{\text{total flux}}
\end{equation}

where the continuum flux (from the {\it apecnoline}$+$power-law fit) and total flux (from the {\it vpshock}$+$power-law fit) are measured within the associated element's bandpass (see Table~\ref{table:lines}). Prior to producing the continuum-subtracted images, we smoothed the derived $F_{\rm elem}$ by taking the average of all adjacent regions. This gave us seven fits files, one for each element (O, Mg, Si, S, Ar, Ca, Fe), with pixel values equal to the smoothed percentage of emission from the given element ($F_{\rm elem}$). We then multiplied the narrow-band images by these fits files to get the continuum-subtracted images shown in Figure~\ref{fig:images}.

We note that in addition to the ejecta, the emission has contributions from shocked interstellar and circumstellar material (ISM, CSM). For example, \cite{hwang12} found CSM-dominated regions in Cas~A along the outer rim of Cas~A, with some in the southwest interior as well. As the periphery of the SNR contributes less flux than the inner regions, it is likely that the emission measure maps (and thus the derived power ratios) are not affected significantly by this CSM contribution.

\subsection{Element Emission Measure Maps} \label{subsec:masscalc}
The final step in our analysis was to convert our continuum-subtracted flux images of each element into emission measure maps:

\begin{equation}
    \rm{Emission \ Measure}= \int n_{\rm{e}}n_{\rm{ion}}dV_{\rm{emit}} \propto \frac{\text{Flux}_{\rm{line}}}{\epsilon(T_e)}  
\end{equation}

\noindent
where $\epsilon(T_e)$ is the emissivity of a line at a given temperature, $n_e$ is electron number density, $n_{\rm{ion}}$ is ion number density, and V$_{\rm{emit}}$ is the emitting volume of the region. The emission measure maps of each element were created by dividing the continuum-subtracted flux images by the emissivity map of each element at the spectrally-fit temperature and ionization timescale.\footnote{The nei emissivity values for each ion at a given temperature were taken from the latest nei emissivity ATOMDB table, `apec\_v3.0.9\_nei\_line.fits'.}

We use the emission measure as a proxy for mass, which holds true under the assumption that the electron number density is constant throughout the remnant.

\subsection{Titanium Data} \label{subsec:Ti}

In addition to the {\it Chandra} images, we also analyzed the 4.6~Ms {\it NuSTAR} $^{44}$Ti (65--70 keV) image of Cas~A, as presented in \cite{grefenstette14, grefenstette17}. We did not perform spatially-resolved continuum subtraction on this narrow-band image; as noted in Section~\ref{sec:data}, we estimate that 80--100\% of the flux in this band is from the radioactive decay line based on fits to the non-thermal continuum presented in \cite{grefenstette15}.

\section{Methods} \label{sec:methods}

We use the power-ratio method (PRM), a multipole expansion technique, to analyze the distribution of elements in Cas~A. This method was employed previously to characterize the X-ray morphology of galaxy clusters \citep{buote95,buote96,jeltema05} and was adapted by \cite{lopez09a} for use on SNRs (e.g., \citealt{lopez09b,lopez11,peters13,me17,stafford18,lopez18}). Using the PRM, we calculate the powers $P_{\rm m}$ of the expansion, which are derived by integrating the magnitude of the $m$-th term over a circle of radius $R$. Then we divide the powers $P_{\rm m}$ by the zeroeth order term $P_{0}$ to normalize with respect to flux. For a more detailed/mathematical description of this method and its application to SNRs, see \cite{lopez09a}. 

\begin{figure*}
\begin{center}
\includegraphics[width=0.67\columnwidth]{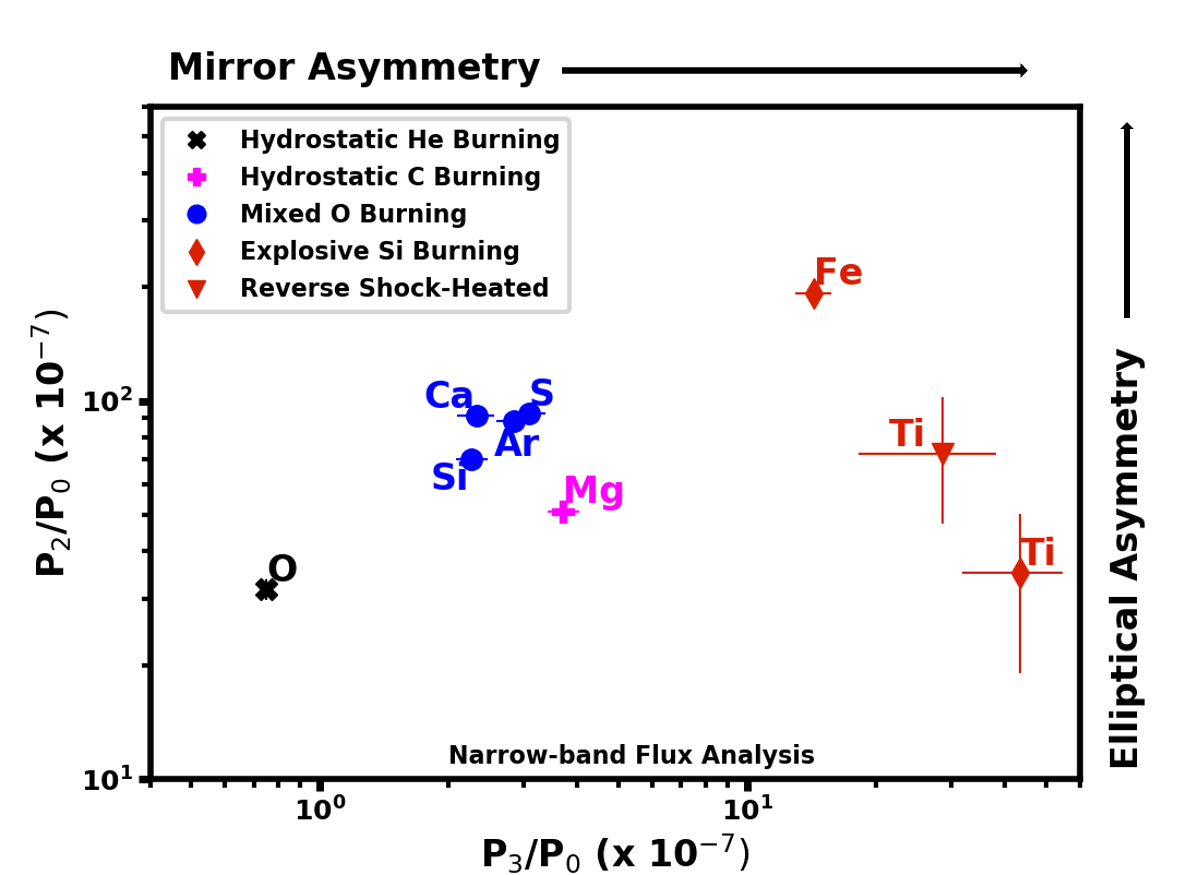}
\includegraphics[width=0.67\columnwidth]{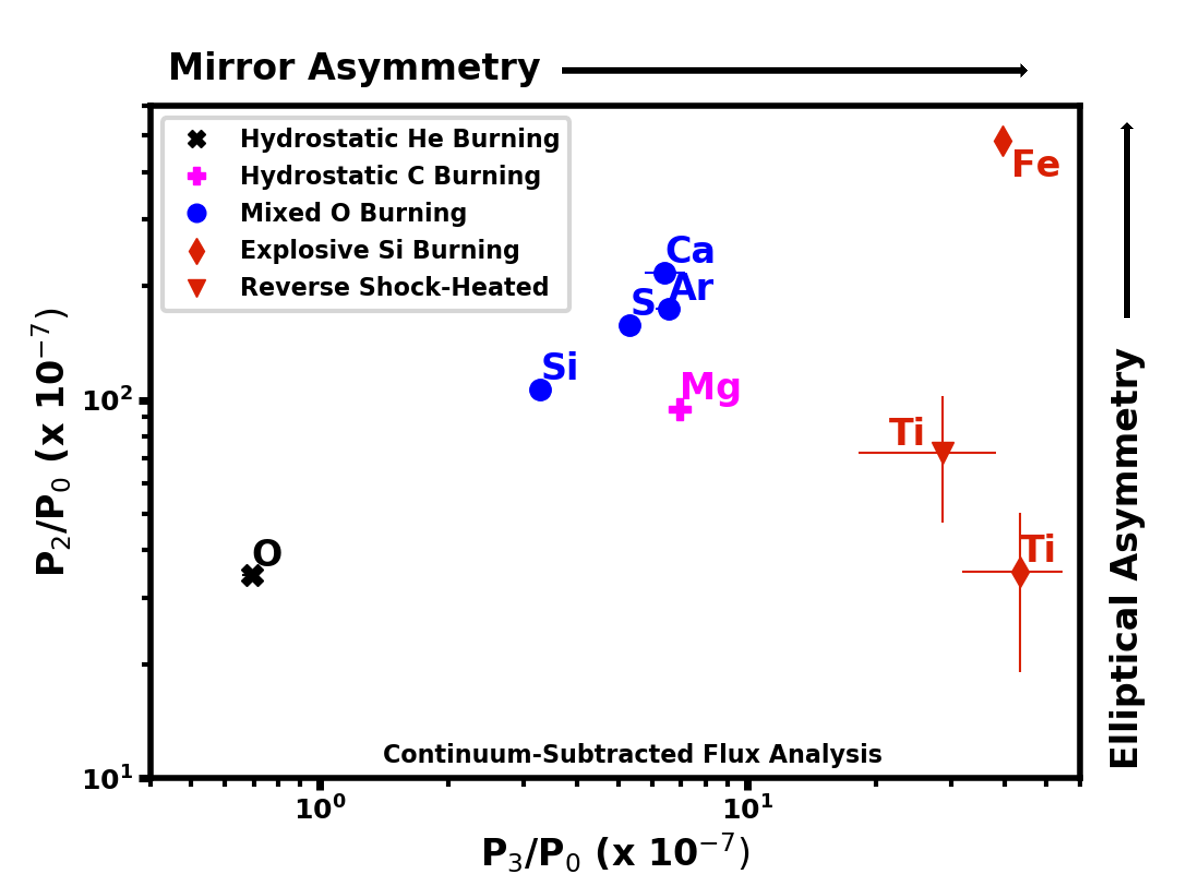}
\includegraphics[width=0.67\columnwidth]{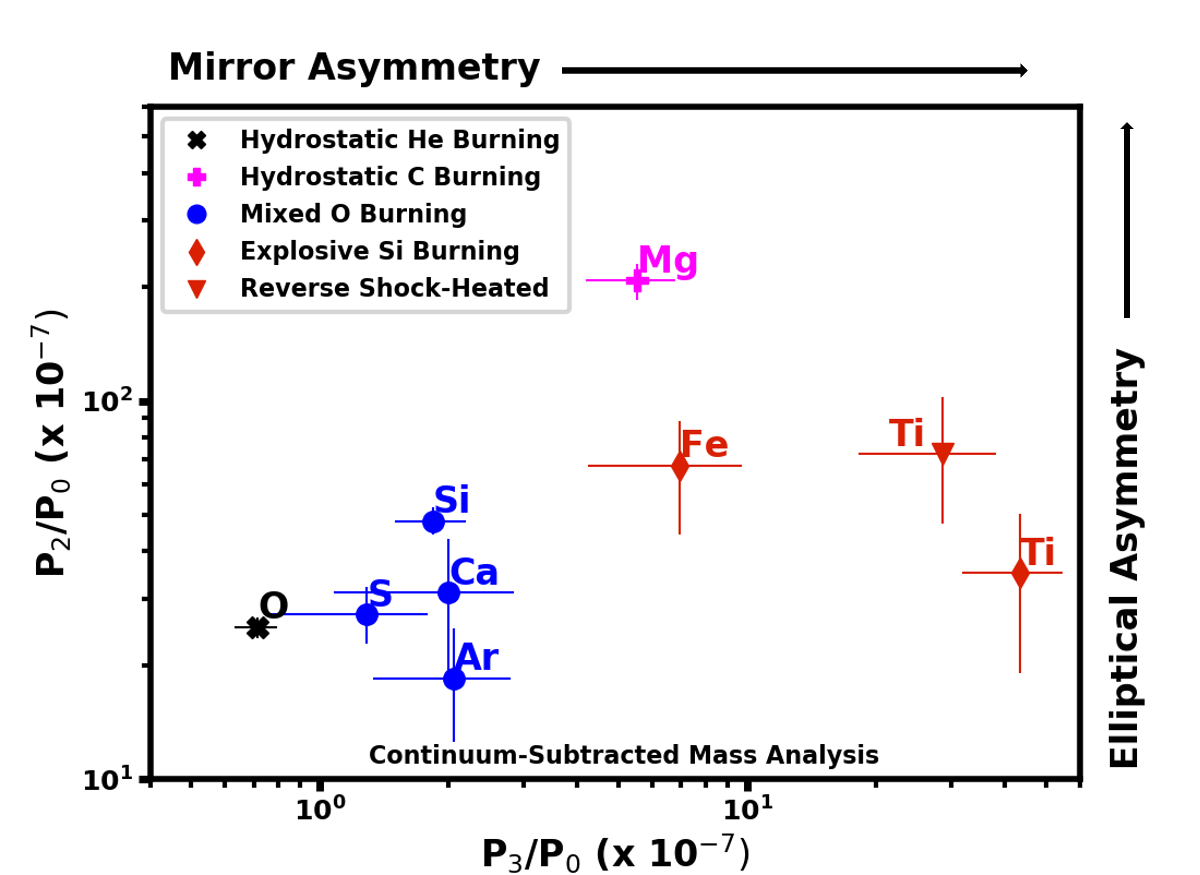}

\end{center}
\caption{The quadrupole power-ratios $P_{2}/P_{0}$ versus the octupole power-ratios $P_{3}/P_{0}$ using the narrow-band images (left), continuum-subtracted images (middle), and the emission measure maps (right). The elements are color-coded by the main burning process that creates them. The $^{44}$Ti data points on both panels are from analysis on the narrow-band Ti image. The ``Reverse Shock-Heated" Ti data point represents the analysis of the image where all Ti flux interior the reverse shock has been set to zero, to better compare the radioactive Ti emission with the emission from the other reverse shock-heated elements (see Section~\ref{sec:results}).}
\label{fig:com}
\end{figure*}

For flux and emission measure maps, we adopt the center-of-emission of each element's image (a proxy for the center-of-mass) as the origin of the multipole expansion for that element (marked as filled blue circles in Figure~\ref{fig:images}). In this case, the dipole power-ratio $P_{1}$/P$_0$ approaches zero, whereas the higher-order moments give details about successively smaller-scale asymmetries. $P_2$/P$_0$ is the quadrupole power-ratio and quantifies the ellipticity/elongation of an extended source. $P_3$/P$_0$ is the octupole power-ratio and is a measure of mirror asymmetry. 

We performed this analysis on all three of our image sets--narrow-band, continuum-subtracted, and emission measure maps--to investigate how closely the narrow-band and continuum-subtracted results matched those of the emission measure maps. For the narrow-band images, uncertainties in the power-ratios are estimated via the Monte Carlo process described in \cite{lopez09a}. The program \emph{AdaptiveBin} \citep{sanders01} is used to bin the SNR into sections of equal total photon counts. Then, the counts in each bin are replaced by a number taken randomly from a Poisson distribution, with the mean equaling the original number of counts. 

For the continuum-subtracted {\it Chandra} images (Figure~\ref{fig:images}), uncertainties in the power-ratios are estimated via a different Monte Carlo process. Each element's fractional contribution to the observed flux $F_{\rm{elem}}$ (values of 0 to 1) has an associated error. This error is greater for weaker emission lines since the fractional contribution of noise is more. We performed a linear extrapolation (python's {\it spline}) on the median of binned errors taken from $F_{\rm{elem}}$ values from 1000 random boxes to create a function that, given a value of $F_{\rm{elem}}$, would output the typical error on that measurement. Similar to the above Monte Carlo method, we then replaced each region's flux fraction with a number taken randomly from a truncated normal distribution, with the mean equaling the original fractional value and the error estimated from the described linear extrapolation.

For the emission measure maps, an additional source of error is the uncertainties in the emissivity. To determine the magnitude of these uncertainties, we produced mock emissivity maps by selecting temperatures and ionization timescales for each region from a truncated normal distribution based on the errors associated with those parameters in the region's spectral fit. Finally, we divided the mock continuum-subtracted images (see above) by the mock emissivity maps.

For analysis using each of the three methods, the process was repeated 100 times to create 100 mock images of each element. We then measured the power-ratios of the 100 images and took the mean of those values as the derived power-ratios plotted in Figure~\ref{fig:com}. We adopt the sixteenth-highest and -lowest values as the confidence limits, chosen to match the 1-$\sigma$ range of a Gaussian distribution.

\section{Results} \label{sec:results}

Figure~\ref{fig:com} shows the quadrupole power-ratio $P_{2}/P_{0}$ versus the octupole power-ratio $P_{3}/P_{0}$ for each element using the narrow-band images, continuum-subtracted images, and emission measure maps. We find a consistent correlation in all of our analyses: the distribution of heavier elements is generally more elliptical and more mirror asymmetric than that of lighter elements. In the following section, we discuss each set of images' assumptions, strengths, and weaknesses. 

The narrow-band images (Fig.~\ref{fig:com}, left) are produced by bandpass cuts corresponding to lines from a single element. As a result, in addition to the line emission, they contain contamination from thermal and non-thermal continuum. Thus, it is necessary to remove the contribution of the continuum to accurately measure asymmetries in line emission. 

The continuum-subtracted images (Fig.~\ref{fig:com}, center) better reflect the spatial distribution of the line emission. Comparing the continuum-subtracted image results to those from the narrow-band images, we find similar trends. The heavier elements (e.g., Fe) are more asymmetric than the lighter elements (e.g., O). For the continuum-subtracted images, the mixed O burning products (Si, S, Ar, and Ca) also follow this trend between them, whereas they are scattered together in the narrow-band plot. Thus, subtraction of the continuum was important to assess the relative asymmetry between these elements from the same burning process (see Section~\ref{subsec:elemdistr} for more details on the burning processes).

The narrow-band and continuum-subtracted images reflect the spatial distribution of emission and not necessarily of mass. Figure~\ref{fig:com} (right) plots the power ratios of the elemental emission measure maps (used as a proxy for mass; described in Section~\ref{subsec:masscalc}) for comparison to the image analyses. We find that oxygen is still the most circular and mirror symmetric, and iron is the most mirror asymmetric. However, the magnitude of the elements' power ratios have decreased compared to those from the image analyses. The error bars on the emission measure map power-ratios are substantially greater than those from the image analyses because the former incorporates several uncertainties from the spectral fitting (e.g., from the temperature and ionization state of the plasma). Thus, it is unclear whether disparities between the imaging and emission measure map analyses arise from the emission measure not being a perfect proxy for mass or from the additional assumptions necessary to compute the emission measures.

\subsection{Elemental Distributions}\label{subsec:elemdistr}

In Figure~\ref{fig:com}, we group the elements by the burning processes that produced them. Oxygen and magnesium are mostly formed prior to the explosion, through hydrostatic core burning of helium and carbon, respectively \citep{woosley02, curtis19}. At the onset of the explosion, these elements are mixed together in an outer convective shell. During the explosion, the innermost section of the oxygen-magnesium-neon shell experiences explosive burning that produces the higher-mass elements \citep{woosley02, curtis19}. 

The remaining O and Mg--located far away from the explosive forces--should exhibit the most symmetric post-explosion distributions compared to the heavier elements. In all of our plots, we find that the O in Cas~A has significantly lower levels of asymmetry than intermediate-mass and heavy elements. By comparison, the location of Mg is inconsistent; in the continuum-subtracted image analysis, it is closer to the power ratios of the intermediate-mass elements, whereas in the emission measure map analysis, its $P_{2}/P_{0}$ is above even that of Fe. We attribute this discrepancy to the presence of non-thermal emission coincident with the regions of highest Mg abundance and the fact that it is difficult to model the X-ray continuum in the soft X-ray bands as there are many emission lines in close proximity.
We find that Fe exhibits the highest levels of ellipticity and mirror asymmetry, running the power-ratio method on both the continuum-subtracted images and the emission measure maps. Our results are broadly consistent with the relative element asymmetries from 3D SN simulations \citep{wongwathanarat13,janka17,muller18}, where O is the most symmetrically distributed, and heavier elements (particularly Fe) are the most asymmetric.

Titanium, an element also formed primarily through explosive silicon burning \citep{woosley02, curtis19}, is predicted to have similar levels of asymmetry as Fe in neutrino-driven explosions \citep{wongwathanarat17}. To test this hypothesis, we measured the power-ratios of the $^{44}$Ti in the narrow-band {\it NuSTAR} image and plot the results in Figure~\ref{fig:com}. While the mirror asymmetry of the Ti is comparable to that of Fe, the ellipticity of the Ti is extremely low. The Ti emission is from radioactive decay, whereas the other elements' line emission is from collisional de-excitation following heating by the reverse shock. As the reverse shock has not fully propagated to the interior of Cas~A \citep{gotthelf01, delaney10}, the ejecta in the SNR center is not hot enough to produce X-ray emission, and thus the elements' symmetry measured from the soft X-ray lines may not reflect the true distribution of those metals.

To better compare the Ti distribution to that of the reverse shock-heated metals, we re-ran the symmetry analysis on a Ti image where the detected emission interior to the revere shock was set to zero. We find that the resulting elliptical asymmetry of Ti increases by a factor of two, whereas the mirror asymmetry decreases by $\sim$30\%, suggesting that the Ti distribution is more consistent with the other elements. We note that $\approx$40\% of the detected Ti in Cas~A is interior to the reverse shock \citep{grefenstette17}, so we caution that this approach ignores a large fraction of the Ti.

The four elements clumped near the center of the power-ratio plot--Si, S, Ar, Ca--are all formed by a mix of hydrostatic and explosive oxygen burning \citep{woosley02, curtis19}. The combination of these two processes, which should result in low and high levels of asymmetry respectively, is likely why the elements have intermediate values of the power-ratios in Figure~\ref{fig:com}. These metals likely have similar degrees of asymmetry as each other because they are formed through the same burning processes.

\begin{figure*}
\begin{center}
\includegraphics[width=\columnwidth]{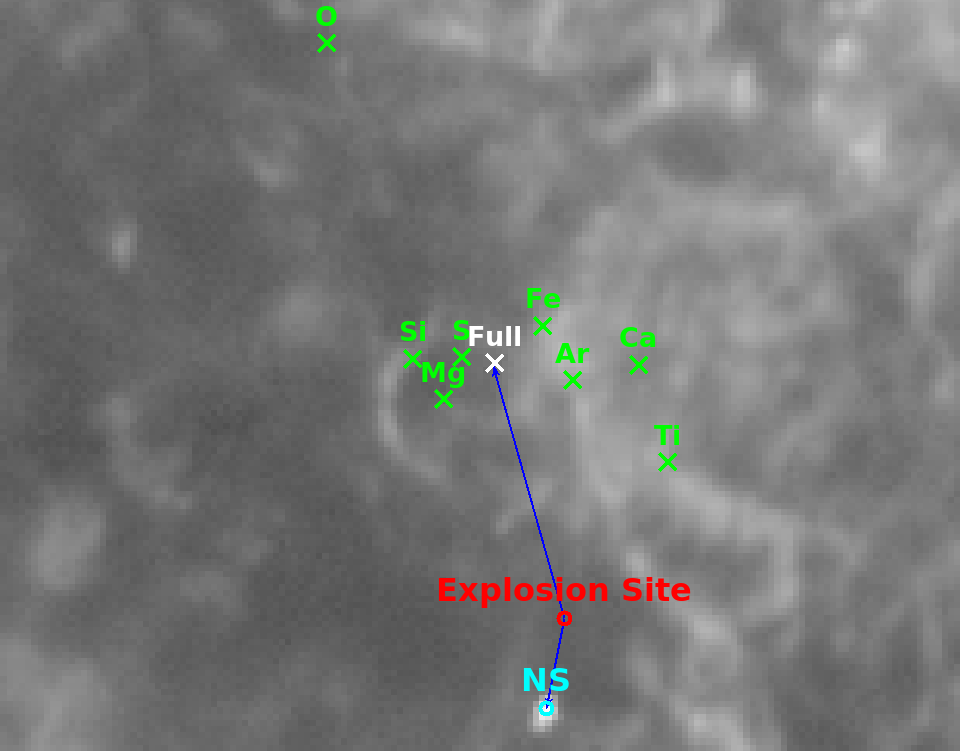}
\includegraphics[width=\columnwidth]{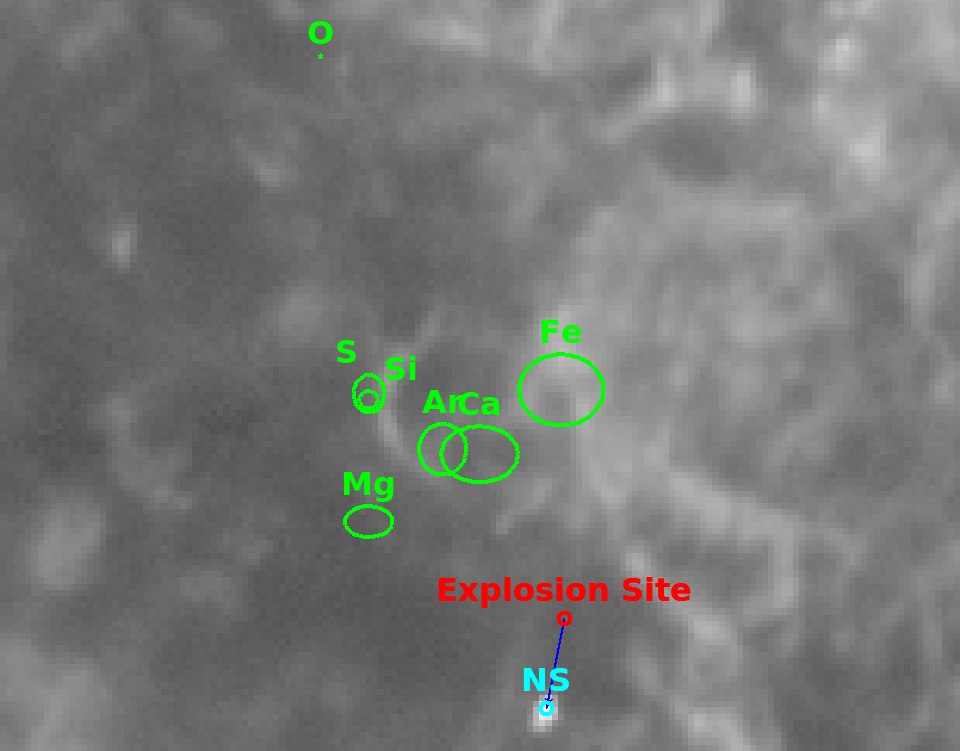}
\end{center}
\caption{Zoomed in images of Cas~A. The left image includes marks of the centers-of-emission for each element (green Xs), the full band (0.5--8.0~keV; white X), the explosion site (red circle), and the current location of the neutron star (cyan circle). The vector from the explosion site to the NS indicates the NS's direction of motion. The vector from the explosion site to full-band center-of-emission indicates the direction of motion of the bulk of ejecta emission, a $\sim$155$^\circ$ angle from the NS direction. The right image includes marks of the emission measure centroid (a proxy for center-of-mass) for each element and the 1-$\sigma$ error bars (green circles), as well as the explosion site and current NS location.}
\label{fig:comlocs}
\end{figure*}

\subsection{Comparison of NS Kicks to Ejecta Distributions}\label{subsec:NSkicks}

We also investigate how the element asymmetries compare to the NS kick direction (see Figure~\ref{fig:comlocs}). Based on the angle between the centers-of-emission for each element and the NS position with respect to the explosion site, the heavier elements (Ar, Ca, Ti, and Fe) are more directly opposed to the NS direction of motion than the lighter elements (O, Mg, Si, and S; see Table~\ref{table:lines}). These results are consistent with recent work by \cite{me17} and \cite{katsuda18} which showed that NSs are preferentially moving in a direction opposed to the bulk of ejecta in several young SNRs, as expected if the NS kick arises from conservation of momentum with the ejecta \citep{wongwathanarat13,muller16,bruenn16, janka17}. 

Recent SN simulations have shown that the ``gravitational tug-boat mechanism"  \citep{wongwathanarat13, janka17} can account for the observed NS velocities of hundreds of km~s$^{-1}$ \citep{lyne94, arz02, hobbs05, faucher06}. In these models, hydrodynamical instabilities in the SN lead to asymmetric mass ejection, accelerating the NS in a direction opposed to the bulk of ejecta by the gravitational forces of the close, slower moving ejecta ``clumps" \citep{scheck06, wongwathanarat13, janka17}. Simulations by \cite{wongwathanarat13} revealed that such gravitational forces from anisotropic ejecta can generate NS recoil velocities of $>$700 km s$^{-1}$, and 2D models by \cite{scheck06} have achieved NS velocities exceeding 1000km$^{-1}$. In this scenario, heavier elements (e.g., Ar, Ca, Ti, Fe) are expelled in a direction opposite to the NS kick, while the lower-mass elements (e.g., O, Mg, Si, S) show a weaker correlation \citep{wongwathanarat13,janka17}. Our results support these predictions, as we find the highest mass elements (Ar, Ca, Ti, Fe) are moving more directly opposed to the NS motion than the intermediate-mass elements.

\subsection{Comparison to Other Studies}\label{subsec:hwang}

Our results are similar to the results of the \cite{hwang12} study on element distributions in Cas~A, with a couple key differences. Consistent with \cite{hwang12}, we find a general N-NE bulk motion of O, Si, S, and Ar, with O having the most extreme total velocity. For these elements, our continuum-subtracted images and emission measure maps match those reported in \cite{hwang12}. However, our maps and bulk motions for Mg and Fe do not match with their results. \cite{hwang12} report a N-NE velocity for Mg, whereas we find that Mg has a N-NE velocity more similar to that of Si and S (using the continuum-subtracted line emission) or that of Ar (using the emission measure maps). Furthermore, we find the bulk Fe is nearly directly north of the explosion site in both the continuum-subtracted images and emission measure maps, making an angle of $\sim165^{\circ}$ with respect to the NS direction of motion. By contrast, \cite{hwang12} find that the bulk Fe (from either explosive silicon burning or from $\alpha$-rich freezeout) is moving at an angle perpendicular to the NS's motion. Consequently, the results of \cite{hwang12} for Mg and Fe are inconsistent with the predictions associated with the gravitational tug-boat mechanism.

We attribute our differing results from \cite{hwang12} to the models used to produce the emission measure maps and to the spectral models we employed. The maps of \cite{hwang12} are best-fit abundances relative to solar, with which they used densities and volumes from their model to estimate mass, whereas we calculated the fluxes and divided by emissivities to calculate emission measure as a proxy for mass. Each method introduces its own errors into the mass calculation. Furthermore, we included a non-thermal component in our spectral models, whereas \cite{hwang12} did not. While most regions of Cas~A are dominated by thermal emission, others have significant non-thermal contributions to the flux \citep{helder08, uchiyama08}. Specifically, we find significant non-thermal emission in the western rim of Cas~A in our maps, matching the non-thermal maps produced by \cite{grefenstette15} using VLA and {\it NuSTAR} images. As this emission is coincident with regions of significantly enhanced abundances of Mg and Fe, it is possible that the maps of \cite{hwang12} are affected by this lack of a non-thermal component in their spectral models.

\cite{katsuda18} also performed detailed spectral analysis of Cas~A, fitting each region by a linear combination of various spectral components: IME-rich ejecta, Fe-rich ejecta, O-rich ejecta, CSM component, and a power-law component. Their IME-rich and O-rich ejecta maps match our results and those of \cite{hwang12}, but their Fe-rich ejecta map does not show the Fe enhancement to the west seen in our maps. In addition, compared to \cite{katsuda18}, we find more power-law flux along the SNR rim and in the west. Our non-thermal results are consistent with the detection of hard ($>$15~keV) X-rays in these locations with {\it NuSTAR} \cite{grefenstette15}.

Recently, \cite{picquenot19} analyzed the {\it Chandra} observations of Cas~A to test their adaption of the general morphological component analysis (GMCA) method for use on X-ray data. The algorithm was developed initially to extract images of the cosmic microwave background \citep{bobin15,bobin16}, and \cite{picquenot19} extended it to X-ray images, which have two spatial dimensions and energy measurements of each photon. Applying GMCA to Cas~A, \cite{picquenot19} were able to disentangle the red-shifted and blue-shifted components of each element in the X-ray spectrum (see their Figure~10). Their resulting maps look similar to our images and emission measure maps of Cas~A, particularly the red-shifted distributions. In a future paper (Picquenot et al., in preparation), they are performing a detailed analysis of Cas~A and exploring the implications regarding the three-dimensional structure of ejecta.

\section{Conclusion} \label{sec:conc}

We have investigated the asymmetries of several elements in Cas~A using X-ray images. We showed that the heaviest elements (Ti, Fe) have more asymmetric morphologies than lighter elements (Mg, Si, S, Ar, Ca), with O the most symmetrically distributed of the elements considered. These results are consistent with predictions from recent, high-fidelity 3D simulations of SN explosions \citep{wongwathanarat13, janka17}. We show that, although the general trend of a linear correlation between elliptical and mirror asymmetry is found using both the continuum-subtracted and the narrow-band images, analysis using continuum-subtracted images is necessary to distinguish the relative asymmetries of elements formed by the same burning process. We find consistent results between analysis on the continuum-subtracted images and emission measure maps created from diving observed flux by emissivity. However, the error bars on the emission measure maps are too uncertain to disentangle the relative levels of asymmetries between elements created via the same burning process.

Though the Ti is as or even more mirror asymmetric as Fe, its observed ellipticity is quite low and does not follow the trends found with the other elements. We attribute this difference to the fact that the hard X-rays from Ti are produced via radioactive decay, whereas the emission from the other elements requires reverse shock-heating. The radioactive Ti is more centrally concentrated than the other elements, since the reverse shock has not propagated to the SNR interior yet. 

Furthermore, we show that the NS kick is most directly opposed to the distribution of the bulk of the heaviest elements, consistent with NS kicks arising from the gravitational tug-boat mechanism \citep{wongwathanarat13, janka17}. 

In the future, application of this analysis to other CC SNRs would be beneficial. A systematic study of element asymmetries in many CC SNRs would reveal the element asymmetries and show the effects that explosion properties (e.g., ejecta mass, explosion energy) have on the metal distributions. This analysis can also be performed on Type Ia SNRs (e.g., Tycho, Kepler, or G1.9$+$0.3) to explore chemical mixing and predictions from different ignition (see e.g., \citealt{badenes06}) and progenitor scenarios (e.g., single- vs double-degenerate).

\acknowledgments

We thank the members of the supernova group in the astronomy department at The Ohio State University, especially Dr. Tuguldur Sukhbold, for their helpful discussions and feedback. We also thank Dr. Brian Grefenstette for sharing the {\it NuSTAR} data and for useful discussions. L.A.L. acknowledges support from the Sophie and Tycho Brahe Visiting Professorship at the Niels Bohr Institute. This work was supported through NSF Astronomy \& Astrophysics Grant AST--1517021. This research made use of the data analysis software: \software{CIAO (v4.7; Fruscione et al. 2006), XSPEC (v12.9.0; Arnaud 1996), Adaptive-Bin (Sanders \& Fabian 2016), ftools (Blackburn 1995)}

\bibliography{CasA_Elem}

\begin{thebibliography}{}
\expandafter\ifx\csname natexlab\endcsname\relax\def\natexlab#1{#1}\fi

\bibitem[{{Alan} {et~al.}(2019){Alan}, {Park}, \& {Bilir}}]{alan19}
{Alan}, N., {Park}, S., \& {Bilir}, S. 2019, \apj, 873, 53

\bibitem[{{Anders} \& {Grevesse}(1989)}]{anders89}
{Anders}, E., \& {Grevesse}, N. 1989, \gca, 53, 197

\bibitem[{{Arnaud}(1996)}]{arnaud96}
{Arnaud}, K.~A. 1996, in Astronomical Society of the Pacific Conference Series,
  Vol. 101, Astronomical Data Analysis Software and Systems V, ed. G.~H.
  {Jacoby} \& J.~{Barnes}, 17

\bibitem[{{Arzoumanian} {et~al.}(2002){Arzoumanian}, {Chernoff}, \&
  {Cordes}}]{arz02}
{Arzoumanian}, Z., {Chernoff}, D.~F., \& {Cordes}, J.~M. 2002, \apj, 568, 289

\bibitem[{{Badenes} {et~al.}(2006){Badenes}, {Borkowski}, {Hughes}, {Hwang}, \&
  {Bravo}}]{badenes06}
{Badenes}, C., {Borkowski}, K.~J., {Hughes}, J.~P., {Hwang}, U., \& {Bravo}, E.
  2006, \apj, 645, 1373

\bibitem[{{Bhalerao} {et~al.}(2019){Bhalerao}, {Park}, {Schenck}, {Post}, \&
  {Hughes}}]{bhalerao19}
{Bhalerao}, J., {Park}, S., {Schenck}, A., {Post}, S., \& {Hughes}, J.~P. 2019,
  \apj, 872, 31

\bibitem[{{Blackburn}(1995)}]{blackburn95}
{Blackburn}, J.~K. 1995, in Astronomical Society of the Pacific Conference
  Series, Vol.~77, Astronomical Data Analysis Software and Systems IV, ed.
  R.~A. {Shaw}, H.~E. {Payne}, \& J.~J.~E. {Hayes}, 367

\bibitem[{{Bleeker} {et~al.}(2001){Bleeker}, {Willingale}, {van der Heyden},
  {Dennerl}, {Kaastra}, {Aschenbach}, \& {Vink}}]{bleeker01}
{Bleeker}, J.~A.~M., {Willingale}, R., {van der Heyden}, K., {et~al.} 2001,
  \aap, 365, L225

\bibitem[{{Bobin} {et~al.}(2015){Bobin}, {Rapin}, {Larue}, \&
  {Starck}}]{bobin15}
{Bobin}, J., {Rapin}, J., {Larue}, A., \& {Starck}, J.-L. 2015, IEEE
  Transactions on Signal Processing, 63, 1199

\bibitem[{{Bobin} {et~al.}(2016){Bobin}, {Sureau}, \& {Starck}}]{bobin16}
{Bobin}, J., {Sureau}, F., \& {Starck}, J.~L. 2016, \aap, 591, A50

\bibitem[{{Bruenn} {et~al.}(2016){Bruenn}, {Lentz}, {Hix}, {Mezzacappa},
  {Harris}, {Messer}, {Endeve}, {Blondin}, {Chertkow}, {Lingerfelt},
  {Marronetti}, \& {Yakunin}}]{bruenn16}
{Bruenn}, S.~W., {Lentz}, E.~J., {Hix}, W.~R., {et~al.} 2016, \apj, 818, 123

\bibitem[{{Buote} \& {Tsai}(1995)}]{buote95}
{Buote}, D.~A., \& {Tsai}, J.~C. 1995, \apj, 452, 522

\bibitem[{{Buote} \& {Tsai}(1996)}]{buote96}
---. 1996, \apj, 458, 27

\bibitem[{{Chen} {et~al.}(2011){Chen}, {Singh}, \& {Cameron}}]{chen11}
{Chen}, J., {Singh}, B., \& {Cameron}, J.~A. 2011, Nuclear Data Sheets, 112,
  2357

\bibitem[{{Chevalier} \& {Kirshner}(1978)}]{chevalier78}
{Chevalier}, R.~A., \& {Kirshner}, R.~P. 1978, \apj, 219, 931

\bibitem[{{Colgate} \& {White}(1966)}]{colgate66}
{Colgate}, S.~A., \& {White}, R.~H. 1966, \apj, 143, 626

\bibitem[{{Curtis} {et~al.}(2019){Curtis}, {Ebinger}, {Fr{\"o}hlich}, {Hempel},
  {Perego}, {Liebend{\"o}rfer}, \& {Thielemann}}]{curtis19}
{Curtis}, S., {Ebinger}, K., {Fr{\"o}hlich}, C., {et~al.} 2019, \apj, 870, 2

\bibitem[{{DeLaney} {et~al.}(2010){DeLaney}, {Rudnick}, {Stage}, {Smith},
  {Isensee}, {Rho}, {Allen}, {Gomez}, {Kozasa}, {Reach}, {Davis}, \&
  {Houck}}]{delaney10}
{DeLaney}, T., {Rudnick}, L., {Stage}, M.~D., {et~al.} 2010, \apj, 725, 2038

\bibitem[{{Faucher-Gigu{\`e}re} \& {Kaspi}(2006)}]{faucher06}
{Faucher-Gigu{\`e}re}, C.-A., \& {Kaspi}, V.~M. 2006, \apj, 643, 332

\bibitem[{{Fesen}(2001)}]{fesen01}
{Fesen}, R.~A. 2001, \apjs, 133, 161

\bibitem[{{Fesen} \& {Milisavljevic}(2016)}]{fesen16}
{Fesen}, R.~A., \& {Milisavljevic}, D. 2016, \apj, 818, 17

\bibitem[{{Fesen} {et~al.}(2006){Fesen}, {Hammell}, {Morse}, {Chevalier},
  {Borkowski}, {Dopita}, {Gerardy}, {Lawrence}, {Raymond}, \& {van den
  Bergh}}]{fesen06}
{Fesen}, R.~A., {Hammell}, M.~C., {Morse}, J., {et~al.} 2006, \apj, 645, 283

\bibitem[{{Foster} {et~al.}(2012){Foster}, {Ji}, {Smith}, \&
  {Brickhouse}}]{foster12}
{Foster}, A.~R., {Ji}, L., {Smith}, R.~K., \& {Brickhouse}, N.~S. 2012, \apj,
  756, 128

\bibitem[{{Gerardy} \& {Fesen}(2001)}]{gerardy01}
{Gerardy}, C.~L., \& {Fesen}, R.~A. 2001, \aj, 121, 2781

\bibitem[{{Gessner} \& {Janka}(2018)}]{gessner18}
{Gessner}, A., \& {Janka}, H.-T. 2018, \apj, 865, 61

\bibitem[{{Gotthelf} {et~al.}(2001){Gotthelf}, {Koralesky}, {Rudnick}, {Jones},
  {Hwang}, \& {Petre}}]{gotthelf01}
{Gotthelf}, E.~V., {Koralesky}, B., {Rudnick}, L., {et~al.} 2001, \apjl, 552,
  L39

\bibitem[{{Grefenstette} {et~al.}(2014){Grefenstette}, {Harrison}, {Boggs},
  {Reynolds}, {Fryer}, {Madsen}, {Wik}, {Zoglauer}, {Ellinger}, {Alexander},
  {An}, {Barret}, {Christensen}, {Craig}, {Forster}, {Giommi}, {Hailey},
  {Hornstrup}, {Kaspi}, {Kitaguchi}, {Koglin}, {Mao}, {Miyasaka}, {Mori},
  {Perri}, {Pivovaroff}, {Puccetti}, {Rana}, {Stern}, {Westergaard}, \&
  {Zhang}}]{grefenstette14}
{Grefenstette}, B.~W., {Harrison}, F.~A., {Boggs}, S.~E., {et~al.} 2014, \nat,
  506, 339

\bibitem[{{Grefenstette} {et~al.}(2015){Grefenstette}, {Reynolds}, {Harrison},
  {Humensky}, {Boggs}, {Fryer}, {DeLaney}, {Madsen}, {Miyasaka}, {Wik},
  {Zoglauer}, {Forster}, {Kitaguchi}, {Lopez}, {Nynka}, {Christensen}, {Craig},
  {Hailey}, {Stern}, \& {Zhang}}]{grefenstette15}
{Grefenstette}, B.~W., {Reynolds}, S.~P., {Harrison}, F.~A., {et~al.} 2015,
  \apj, 802, 15

\bibitem[{{Grefenstette} {et~al.}(2017){Grefenstette}, {Fryer}, {Harrison},
  {Boggs}, {DeLaney}, {Laming}, {Reynolds}, {Alexander}, {Barret},
  {Christensen}, {Craig}, {Forster}, {Giommi}, {Hailey}, {Hornstrup},
  {Kitaguchi}, {Koglin}, {Lopez}, {Mao}, {Madsen}, {Miyasaka}, {Mori}, {Perri},
  {Pivovaroff}, {Puccetti}, {Rana}, {Stern}, {Westergaard}, {Wik}, {Zhang}, \&
  {Zoglauer}}]{grefenstette17}
{Grefenstette}, B.~W., {Fryer}, C.~L., {Harrison}, F.~A., {et~al.} 2017, \apj,
  834, 19

\bibitem[{{Helder} \& {Vink}(2008)}]{helder08}
{Helder}, E.~A., \& {Vink}, J. 2008, \apj, 686, 1094

\bibitem[{{Hobbs} {et~al.}(2005){Hobbs}, {Lorimer}, {Lyne}, \&
  {Kramer}}]{hobbs05}
{Hobbs}, G., {Lorimer}, D.~R., {Lyne}, A.~G., \& {Kramer}, M. 2005, \mnras,
  360, 974

\bibitem[{{Holland-Ashford} {et~al.}(2017){Holland-Ashford}, {Lopez},
  {Auchettl}, {Temim}, \& {Ramirez-Ruiz}}]{me17}
{Holland-Ashford}, T., {Lopez}, L.~A., {Auchettl}, K., {Temim}, T., \&
  {Ramirez-Ruiz}, E. 2017, \apj, 844, 84

\bibitem[{{Hughes} {et~al.}(2000){Hughes}, {Rakowski}, {Burrows}, \&
  {Slane}}]{hughes00}
{Hughes}, J.~P., {Rakowski}, C.~E., {Burrows}, D.~N., \& {Slane}, P.~O. 2000,
  \apjl, 528, L109

\bibitem[{{Hwang} {et~al.}(2000){Hwang}, {Holt}, \& {Petre}}]{hwang00}
{Hwang}, U., {Holt}, S.~S., \& {Petre}, R. 2000, \apjl, 537, L119

\bibitem[{{Hwang} \& {Laming}(2003)}]{hwang03}
{Hwang}, U., \& {Laming}, J.~M. 2003, \apj, 597, 362

\bibitem[{{Hwang} \& {Laming}(2012)}]{hwang12}
---. 2012, \apj, 746, 130

\bibitem[{{Hwang} {et~al.}(2008){Hwang}, {Petre}, \& {Flanagan}}]{hwang08}
{Hwang}, U., {Petre}, R., \& {Flanagan}, K.~A. 2008, \apj, 676, 378

\bibitem[{{Janka}(2017)}]{janka17}
{Janka}, H.-T. 2017, \apj, 837, 84

\bibitem[{{Janka} {et~al.}(2017){Janka}, {Gabler}, \&
  {Wongwathanarat}}]{janka17b}
{Janka}, H.-T., {Gabler}, M., \& {Wongwathanarat}, A. 2017, in IAU Symposium,
  Vol. 331, Supernova 1987A:30 years later - Cosmic Rays and Nuclei from
  Supernovae and their Aftermaths, ed. A.~{Marcowith}, M.~{Renaud},
  G.~{Dubner}, A.~{Ray}, \& A.~{Bykov}, 148--156

\bibitem[{{Janka} {et~al.}(2016){Janka}, {Melson}, \& {Summa}}]{janka16}
{Janka}, H.-T., {Melson}, T., \& {Summa}, A. 2016, Annual Review of Nuclear and
  Particle Science, 66, 341

\bibitem[{{Jeltema} {et~al.}(2005){Jeltema}, {Canizares}, {Bautz}, \&
  {Buote}}]{jeltema05}
{Jeltema}, T.~E., {Canizares}, C.~R., {Bautz}, M.~W., \& {Buote}, D.~A. 2005,
  \apj, 624, 606

\bibitem[{{Katsuda} {et~al.}(2018){Katsuda}, {Morii}, {Janka},
  {Wongwathanarat}, {Nakamura}, {Kotake}, {Mori}, {M{\"u}ller}, {Takiwaki},
  {Tanaka}, {Tominaga}, \& {Tsunemi}}]{katsuda18}
{Katsuda}, S., {Morii}, M., {Janka}, H.-T., {et~al.} 2018, \apj, 856, 18

\bibitem[{{Koo} {et~al.}(2018){Koo}, {Kim}, {Lee}, {Raymond}, {Lee}, {Yoon}, \&
  {Moon}}]{bonchul18}
{Koo}, B.-C., {Kim}, H.-J., {Lee}, Y.-H., {et~al.} 2018, \apj, 866, 139

\bibitem[{{Laming} \& {Hwang}(2003)}]{laming03}
{Laming}, J.~M., \& {Hwang}, U. 2003, \apj, 597, 347

\bibitem[{{Lopez} \& {Fesen}(2018)}]{lopez18}
{Lopez}, L.~A., \& {Fesen}, R.~A. 2018, \ssr, 214, 44

\bibitem[{{Lopez} {et~al.}(2009b){Lopez}, {Ramirez-Ruiz}, {Badenes},
  {Huppenkothen}, {Jeltema}, \& {Pooley}}]{lopez09b}
{Lopez}, L.~A., {Ramirez-Ruiz}, E., {Badenes}, C., {et~al.} 2009b, \apjl, 706,
  L106

\bibitem[{{Lopez} {et~al.}(2011){Lopez}, {Ramirez-Ruiz}, {Huppenkothen},
  {Badenes}, \& {Pooley}}]{lopez11}
{Lopez}, L.~A., {Ramirez-Ruiz}, E., {Huppenkothen}, D., {Badenes}, C., \&
  {Pooley}, D.~A. 2011, \apj, 732, 114

\bibitem[{{Lopez} {et~al.}(2009a){Lopez}, {Ramirez-Ruiz}, {Pooley}, \&
  {Jeltema}}]{lopez09a}
{Lopez}, L.~A., {Ramirez-Ruiz}, E., {Pooley}, D.~A., \& {Jeltema}, T.~E. 2009a,
  \apj, 691, 875

\bibitem[{{Luna} {et~al.}(2016){Luna}, {Smith}, {Dubner}, {Giacani}, \&
  {Castelletti}}]{luna16}
{Luna}, G.~J.~M., {Smith}, M.~J.~S., {Dubner}, G., {Giacani}, E., \&
  {Castelletti}, G. 2016, \aap, 590, A70

\bibitem[{{Lyne} \& {Lorimer}(1994)}]{lyne94}
{Lyne}, A.~G., \& {Lorimer}, D.~R. 1994, \nat, 369, 127

\bibitem[{{Markert} {et~al.}(1983){Markert}, {Canizares}, {Clark}, \&
  {Winkler}}]{markert82}
{Markert}, T.~H., {Canizares}, C.~R., {Clark}, G.~W., \& {Winkler}, P.~F. 1983,
  \apj, 268, 134

\bibitem[{{Milisavljevic} \& {Fesen}(2013)}]{milisavljevic13}
{Milisavljevic}, D., \& {Fesen}, R.~A. 2013, \apj, 772, 134

\bibitem[{{Milisavljevic} \& {Fesen}(2015)}]{milisavljevic15}
---. 2015, Science, 347, 526

\bibitem[{{M{\"u}ller}(2016)}]{muller16}
{M{\"u}ller}, B. 2016, Publications of the Astronomical Society of Australia,
  33, e048

\bibitem[{{M{\"u}ller} {et~al.}(2017){M{\"u}ller}, {Melson}, {Heger}, \&
  {Janka}}]{muller17a}
{M{\"u}ller}, B., {Melson}, T., {Heger}, A., \& {Janka}, H.-T. 2017, \mnras,
  472, 491

\bibitem[{{M{\"u}ller} {et~al.}(2018){M{\"u}ller}, {Tauris}, {Heger},
  {Banerjee}, {Qian}, {Powell}, {Chan}, {Gay}, \& {Langer}}]{muller18}
{M{\"u}ller}, B., {Tauris}, T.~M., {Heger}, A., {et~al.} 2018, ArXiv e-prints,
  arXiv:1811.05483

\bibitem[{{Orlando} {et~al.}(2016){Orlando}, {Miceli}, {Pumo}, \&
  {Bocchino}}]{orlando16}
{Orlando}, S., {Miceli}, M., {Pumo}, M.~L., \& {Bocchino}, F. 2016, \apj, 822,
  22

\bibitem[{{Park} {et~al.}(2002){Park}, {Roming}, {Hughes}, {Slane}, {Burrows},
  {Garmire}, \& {Nousek}}]{park02}
{Park}, S., {Roming}, P. W.~A., {Hughes}, J.~P., {et~al.} 2002, \apjl, 564, L39

\bibitem[{{Peters} {et~al.}(2013){Peters}, {Lopez}, {Ramirez-Ruiz}, {Stassun},
  \& {Figueroa-Feliciano}}]{peters13}
{Peters}, C.~L., {Lopez}, L.~A., {Ramirez-Ruiz}, E., {Stassun}, K.~G., \&
  {Figueroa-Feliciano}, E. 2013, \apjl, 771, L38

\bibitem[{{Picquenot} {et~al.}(2019){Picquenot}, {Acero}, {Bobin}, {Maggi},
  {Ballet}, \& {Pratt}}]{picquenot19}
{Picquenot}, A., {Acero}, F., {Bobin}, J., {et~al.} 2019, \aap, 627, A139

\bibitem[{{Reed} {et~al.}(1995){Reed}, {Hester}, {Fabian}, \&
  {Winkler}}]{reed95}
{Reed}, J.~E., {Hester}, J.~J., {Fabian}, A.~C., \& {Winkler}, P.~F. 1995,
  \apj, 440, 706

\bibitem[{{Rest} {et~al.}(2011){Rest}, {Foley}, {Sinnott}, {Welch}, {Badenes},
  {Filippenko}, {Bergmann}, {Bhatti}, {Blondin}, {Challis}, {Damke}, {Finley},
  {Huber}, {Kasen}, {Kirshner}, {Matheson}, {Mazzali}, {Minniti}, {Nakajima},
  {Narayan}, {Olsen}, {Sauer}, {Smith}, \& {Suntzeff}}]{rest11}
{Rest}, A., {Foley}, R.~J., {Sinnott}, B., {et~al.} 2011, \apj, 732, 3

\bibitem[{{Rutherford} {et~al.}(2013){Rutherford}, {Dewey},
  {Figueroa-Feliciano}, {Heine}, {Bastien}, {Sato}, \&
  {Canizares}}]{rutherford13}
{Rutherford}, J., {Dewey}, D., {Figueroa-Feliciano}, E., {et~al.} 2013, \apj,
  769, 64

\bibitem[{{Sanders} \& {Fabian}(2016)}]{sanders01}
{Sanders}, J.~S., \& {Fabian}, A.~C. 2016, {AdaptiveBin: Adaptive Binning},
  Astrophysics Source Code Library, ascl:1609.024

\bibitem[{{Scheck} {et~al.}(2006){Scheck}, {Kifonidis}, {Janka}, \&
  {M{\"u}ller}}]{scheck06}
{Scheck}, L., {Kifonidis}, K., {Janka}, H.-T., \& {M{\"u}ller}, E. 2006, \aap,
  457, 963

\bibitem[{{Smith} {et~al.}(2001){Smith}, {Brickhouse}, {Liedahl}, \&
  {Raymond}}]{smith01}
{Smith}, R.~K., {Brickhouse}, N.~S., {Liedahl}, D.~A., \& {Raymond}, J.~C.
  2001, \apjl, 556, L91

\bibitem[{{Smith} \& {Hughes}(2010)}]{smith10}
{Smith}, R.~K., \& {Hughes}, J.~P. 2010, \apj, 718, 583

\bibitem[{{Stafford} {et~al.}(2018){Stafford}, {Lopez}, {Auchettl}, \&
  {Holland-Ashford}}]{stafford18}
{Stafford}, J.~N., {Lopez}, L.~A., {Auchettl}, K., \& {Holland-Ashford}, T.
  2018, arXiv e-prints, arXiv:1808.08234

\bibitem[{{Summa} {et~al.}(2018){Summa}, {Janka}, {Melson}, \&
  {Marek}}]{summa18}
{Summa}, A., {Janka}, H.-T., {Melson}, T., \& {Marek}, A. 2018, \apj, 852, 28

\bibitem[{{Tananbaum}(1999)}]{tananbaum99}
{Tananbaum}, H. 1999, \iaucirc, 7246

\bibitem[{{Thorstensen} {et~al.}(2001){Thorstensen}, {Fesen}, \& {van den
  Bergh}}]{thorstensen01}
{Thorstensen}, J.~R., {Fesen}, R.~A., \& {van den Bergh}, S. 2001, \aj, 122,
  297

\bibitem[{{Uchiyama} \& {Aharonian}(2008)}]{uchiyama08}
{Uchiyama}, Y., \& {Aharonian}, F.~A. 2008, \apjl, 677, L105

\bibitem[{{Utrobin} {et~al.}(2017){Utrobin}, {Wongwathanarat}, {Janka}, \&
  {M{\"u}ller}}]{utrobin17}
{Utrobin}, V.~P., {Wongwathanarat}, A., {Janka}, H.-T., \& {M{\"u}ller}, E.
  2017, \apj, 846, 37

\bibitem[{{Vink}(2012)}]{vink12}
{Vink}, J. 2012, \aapr, 20, 49

\bibitem[{{Vink} {et~al.}(1996){Vink}, {Kaastra}, \& {Bleeker}}]{vink96}
{Vink}, J., {Kaastra}, J.~S., \& {Bleeker}, J.~A.~M. 1996, \aap, 307, L41

\bibitem[{{Vink} \& {Laming}(2003)}]{vink03}
{Vink}, J., \& {Laming}, J.~M. 2003, \apj, 584, 758

\bibitem[{{Weisskopf} \& {Hughes}(2006)}]{weiss06b}
{Weisskopf}, M.~C., \& {Hughes}, J.~P. 2006, {Six Years of Chandra Observations
  of Supernova Remnants}, ed. J.~W. {Mason}, 55

\bibitem[{{Wongwathanarat} {et~al.}(2013){Wongwathanarat}, {Janka}, \&
  {M{\"u}ller}}]{wongwathanarat13}
{Wongwathanarat}, A., {Janka}, H.-T., \& {M{\"u}ller}, E. 2013, \aap, 552, A126

\bibitem[{{Wongwathanarat} {et~al.}(2017){Wongwathanarat}, {Janka},
  {M{\"u}ller}, {Pllumbi}, \& {Wanajo}}]{wongwathanarat17}
{Wongwathanarat}, A., {Janka}, H.-T., {M{\"u}ller}, E., {Pllumbi}, E., \&
  {Wanajo}, S. 2017, \apj, 842, 13

\bibitem[{{Woosley} {et~al.}(2002){Woosley}, {Heger}, \& {Weaver}}]{woosley02}
{Woosley}, S.~E., {Heger}, A., \& {Weaver}, T.~A. 2002, Reviews of Modern
  Physics, 74, 1015

\end{thebibliography}

\end{document}